**Graphical Abstract**

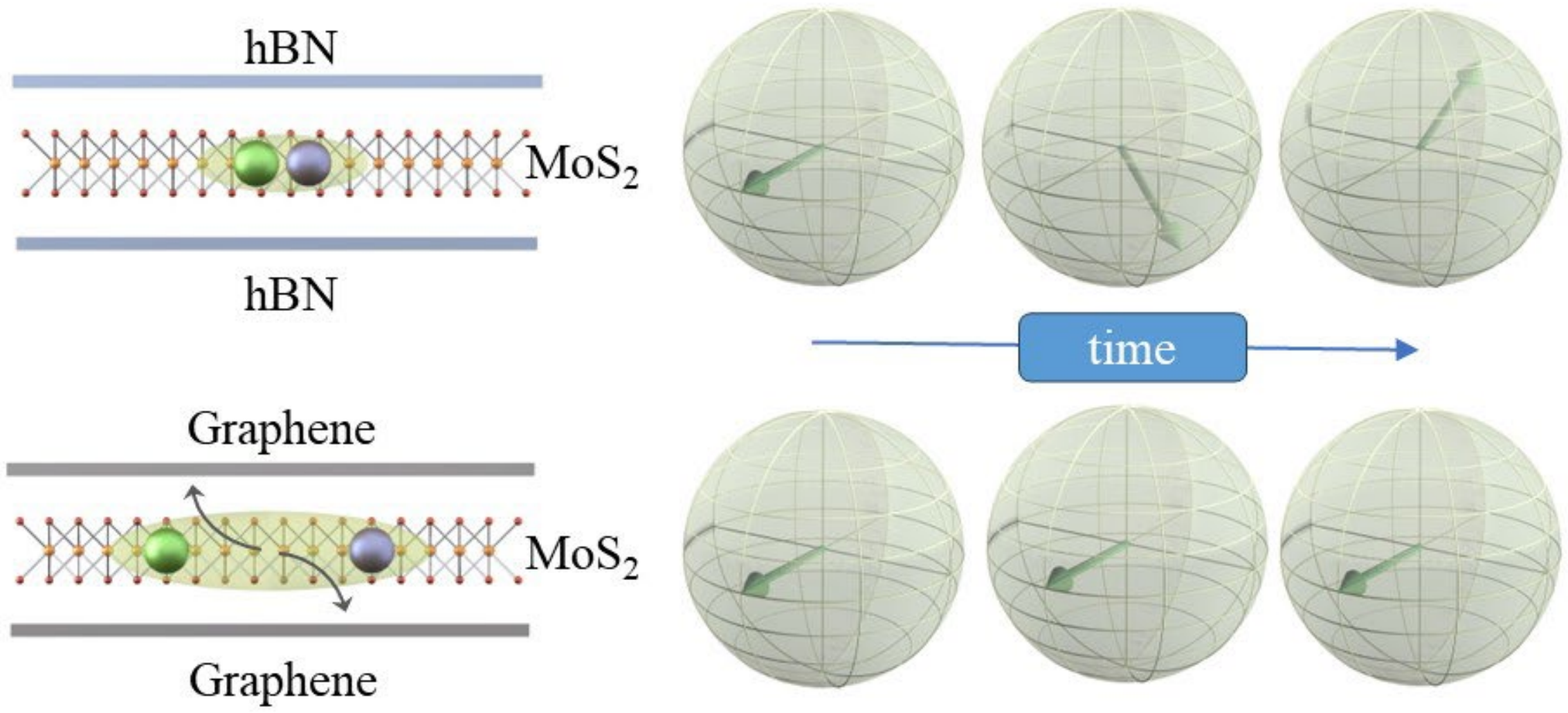

**Short Summary**

A demonstration of the excitons remaining valley coherent during the entire radiative recombination time in few-layer graphene capped monolayer $MoS_2$.

# Observation of ~100% valley-coherent excitons in monolayer $MoS_2$ through giant enhancement of valley coherence time

Garima Gupta[1], Kenji Watanabe[2], Takashi Taniguchi[3], and Kausik Majumdar[1,*]

[1]*Department of Electrical Communication Engineering, Indian Institute of Science, Bangalore 560012, India*

[2] *Research Center for Functional Materials, National Institute for Materials Science, 1-1 Namiki, Tsukuba 305-044, Japan*

[3] *International Center for Materials Nanoarchitectonics, National Institute for Materials Science, 1-1 Namiki, Tsukuba 305-044, Japan*

**corresponding author, E-mail: kausikm@iisc.ac.in*

**Abstract:** In monolayer transition metal dichalcogenide semiconductors, valley coherence degrades rapidly due to a combination of fast scattering and inter-valley exchange interaction. This leads to a sub-picosecond valley coherence time, making coherent manipulation of exciton a highly challenging task. Using monolayer $MoS_2$ sandwiched between top and bottom graphene, here we demonstrate fully valley coherent excitons by observing ~100% degree of linear polarization in steady state photoluminescence. This is achieved in this unique design through a combined effect of (a) suppression in exchange interaction due to enhanced dielectric screening, (b) reduction in exciton lifetime due to a fast inter-layer transfer to graphene, and (c) operating in the motional narrowing regime. We disentangle the role of the key parameters affecting valley coherence by using a combination of calculation (solutions of Bethe-Salpeter and Maialle-Silva-Sham equations) and a careful choice of design of experiments using four different stacks with systematic variation of screening and exciton lifetime. To the best of our knowledge, this is the first report in which the excitons are found to be valley coherent in the entire lifetime in monolayer semiconductors, allowing optical readout of valley coherence possible.

The bound state of an electron and a hole, an exciton, is a superposition of the conduction and valence band states in the $\mathbf{K}$ and $\mathbf{K'}$ valleys in monolayer transition metal dichalcogenides (TMDs)[1–3]. $\mathbf{K}$ and $\mathbf{K'}$ valley excitons are selectively generated by circularly polarized light excitation of opposite helicities[4–7]. On linearly polarized excitation, a hybrid $\mathbf{K}-\mathbf{K'}$ exciton is generated in a state of valley coherence[8,9]. However, valley coherence degrades rapidly due to a combined effect of fast scattering and inter-valley exchange[10–12]. The reported values of valley coherence time lie in the range of $98 - 520$ fs[12–17], much shorter than the exciton radiative lifetime of $\sim 1$ ps[18–20]. This makes optical read out of strong exciton valley coherence a highly challenging task. To be able to use these coherent excitons as a qubit for quantum information processing, a longer valley coherence time is desirable to perform any manipulation on it. Any technique[21] that enhances this valley coherence time significantly is thus of high scientific importance.

Here we demonstrate a $100\%$ degree of linear polarization (DOLP) in photoluminescence (PL) peak of $A_{1s}$ exciton in a monolayer of $MoS_2$ encapsulated with few-layer-graphene (FLG) at the top and bottom. Such a complete retention of the generated valley coherence in steady-state PL implies the achievement of a large valley coherence time, the measured value of which is only limited by the lifetime of the exciton. This suggests that the valley coherence time has been significantly enhanced as compared to the reported values to date[12–16].

Depending on the linear polarization direction of the excitation light, the excitons are generated at specific center-of-mass momentum ($\mathbf{Q}$) values [where $\mathbf{Q} = \mathbf{k_e} + \mathbf{k_h}$, $\mathbf{k_e}(\mathbf{k_h})$ denoting the electron (hole) crystal momentum] in the exciton band at time $t = 0$ (Figure 1). During its lifetime, the exciton undergoes scattering and exchange interaction, which, coupled together, degrades the valley coherence. The polarization state can be represented by the pseudospin vector $\mathbf{S}$ in the Bloch sphere. At $t = 0$, the direction of $\mathbf{S}$ is parallel to the exchange-induced magnetic field (denoted by the precession frequency $\mathbf{\Omega}$). Considering $x-$polarized excitation, the system is generated in a pure state represented by $S_x = 1, S_y = 0, S_z = 0$. When excitons scatter to different $\mathbf{Q}$ values, the pseudospin precesses on experiencing a finite torque around $\mathbf{\Omega}(\mathbf{Q})$ due to which it becomes a mixed state represented by a density matrix operator $\rho$. The $\mathbf{Q}$ - space and the Bloch sphere representations of this mechanism are shown in Figure 1a-b. On decoupling the density matrix in terms of the number trace and traceless matrix $\mathbf{S}.\boldsymbol{\sigma}$ (where $\boldsymbol{\sigma}$ denotes the Pauli matrices), the overall dynamics of the valley pseudospin as described by the Maialle-Silva-Sham (MSS) mechanism (see Supplementary Note 1 for proof) is given by[10]:

$$\frac{d\mathbf{S}(\mathbf{Q})}{dt} = \mathbf{\Omega}(\mathbf{Q}) \times \mathbf{S}(\mathbf{Q}) + \sum_{\mathbf{Q}'} \mathrm{W}_{\mathbf{QQ}'} \left[\mathbf{S}(\mathbf{Q}') - \mathbf{S}(\mathbf{Q})\right] - \frac{1}{\tau}\mathbf{S}(\mathbf{Q}) + \mathbf{G} \qquad (1)$$

Here $\mathrm{W}_{\mathbf{QQ}'}$ is the rate of any generic momentum scattering mechanism, e.g., exciton-impurity and exciton-phonon scattering. $\tau$ is the net exciton lifetime given by ${}^{1}/_{\tau} = {}^{1}/_{\tau_r} + {}^{1}/_{\tau_{nr}} + {}^{1}/_{\tau_{filter}}$, where $\tau_r, \tau_{nr}$, and $\tau_{filter}$ are the radiative, non-radiative, and filtering timescale. Filtering is a non-radiative process where the excitons are scattered out of the light cone, e.g., scattering to lower energy states, interlayer transfer to graphene, etc., in which case, the light collection is limited to $t \leq \tau_{filter}$. **G** represents the exciton generation rate. On recombination, the DOLP of this mixed state is given by $\langle S_x \rangle$, averaged over **Q** values within the light cone (see Supplementary Note 2 for proof).

The possible ways to improve the valley coherence time are: (a) by minimizing scattering ($\mathrm{W}_{\mathbf{QQ}'}$) inside the light cone such that **S** does not accumulate random phase by precessing around $\mathbf{\Omega}(\mathbf{Q})$, or by enhancing $\mathrm{W}_{\mathbf{QQ}'}$ such that the whole operation is pushed towards the motional narrowing regime (simulation results in Figure 1c); (b) by screening the electron-hole interaction which results in reduced exchange interaction, and in turn a suppressed $\mathbf{\Omega}$ (middle panel of Figure 1d). However, a side-effect of the enhanced screening is an increment in the exciton lifetime due to a reduction in the binding energy. This can be overcome by (c) reducing $\tau$ by introducing a fast-filtering mechanism[21–23] (bottom panel of Figure 1d).

To understand the interplay among these factors systematically, we prepare four different stacks of monolayer $MoS_2$ combined with hBN and FLG, which are: (1) hBN-$MoS_2$-hBN (HMH), (2) FLG-hBN-$MoS_2$-hBN-FLG (GHMHG), (3) $MoS_2$-FLG-hBN (MGH) and (4) FLG-$MoS_2$-FLG (GMG) (see Methods). We obtain an exciton DOLP of 44.5 $(\pm 10)\%$ in the HMH stack, 37 $(\pm 9)\%$ in the GHMHG stack, 77 $(\pm 5)\%$ in the MGH stack, and 96 $(\pm 6)\%$ in the GMG stack on 633 nm near resonant laser excitation at 5 K. Linear polarization-resolved representative PL spectra in Figure 2a-d and the bar diagram (with error bars) in Figure 2e compare the DOLP numbers in all the four stacks (more spectra in Supplementary Figure 3-6). Interestingly, there are several spots where we observe $\sim$100% DOLP in the GMG stack (Figure 2d and Supplementary Figure 6). In Supplementary Figure 7a, we show similar results of ~100% DOLP obtained from few-layer graphene encapsulated monolayer $WS_2$ (GWG) stack. We also perform polarization dependent time-resolved photoluminescence (TRPL) measurement and obtain a peak DOLP of 97.6% in the GWG stack (see Supplementary Figure 7b).

We would like to highlight some additional observations on the GMG stack before the main analysis begins: (1) As a result of FLG encapsulation, the PL spectra predominantly consist of the clean $A_{1s}$ exciton peak[22]. The spectra of FLG encapsulated monolayer $MoS_2$, $MoSe_2$, and $WS_2$ on 532 nm excitation are shown in Figure 2f, clearly indicating suppression of spurious peaks from defect-bound excitons and other excitonic complexes. We also observe a clear $A_{2s}$ peak located at 44 (32) meV higher than the $A_{1s}$ peak in $MoS_2$ ($MoSe_2$ and $WS_2$) due to enhanced screening[24]. (2) We also get a very high degree of circular polarization (DOCP) of 81.6 ($\pm 2$) % in the GMG stack, much larger compared to the 20.5 ($\pm 9$) % DOCP in the HMH stack (Figure 2g-h and more spectra in Supplementary Figure 8-9). The in-plane nature of $\mathbf{\Omega}$ explains this observation that DOCP is smaller than DOLP for 2D excitons, which is consistent with previous reports[10,25] (see Supplementary Note 3). This indicates that starting with a linear polarization (that is, on the equator of the Bloch sphere) is the most favourable scenario to maintain valley coherence compared with any other (elliptical) polarization. (3) The initial and the final state in the Raman scattering process coincides with the $A_{2s}$ and the $A_{1s}$ exciton level, respectively, on 633 nm excitation at 5 K in the GMG stack. This dual resonance enhances the intensity of the Raman peaks significantly (represented by the dashed lines in Figure 2d and 2h) and enables the observation of other less commonly observed modes distinctly (Supplementary Figure 10). The fact that the excitation laser is resonant with the 2s state in the GMG stack and we observe almost fully-coherent 1s excitonic emission from the stack, it is likely that the generated 2s excitons relax through polarization preserving processes, such as, dipole-coupled radiative transition (2s→2p→1s).

In order to establish the different degrees of screening in the stacks, we plot the PL spectra for the HMH, GHMHG, and the GMG stack obtained from 532 nm excitation in Figure 3a-c. The $A_{2s} - A_{1s}$ energy separation obtained is 144.5 meV in the HMH stack, which reduces to 60 and 45 meV in the GHMHG and GMG stack, respectively. To get an estimate of the $A_{1s}$ exciton binding energy change (Figure 3d), we obtain the continuum of the exciton energy spectrum by numerically solving the Bethe-Salpeter equation[26] using a two-band Hamiltonian. In the calculation, the parameters are fitted such that the experimentally obtained $A_{2s} - A_{1s}$ energy separation matches with the calculated one (Supplementary Note 4 and Supplementary Figure 2). The calculated $A_{1s}$ exciton binding energy is 379 meV in the HMH system, which reduces to 162.5 and 122 meV in the GHMHG and the GMG stacks respectively due to graphene induced screening.

One immediate consequence of such a screening is the reduction in the inter-valley exchange interaction[25]. The exchange interaction is composed of two components – the short-range part and the long-range part. The short-range component is zero at $\mathbf{Q} = \mathbf{0}$ due to the three-fold rotational symmetry condition and is negligible at higher $\mathbf{Q}$ values. The long-range part is given by[27]:

$$J_{\mathbf{Q}}^{LR} \propto -\frac{|\sum_{\mathbf{k}} \psi(\mathbf{k})|^2}{E_g^2} V(\mathbf{Q})|\mathbf{Q}|^2 \quad (2)$$

Here $\frac{|\sum_{\mathbf{k}} \psi(\mathbf{k})|^2}{A} = |\psi(r_{eh} = 0)|^2$ is the electron-hole wavefunction overlap at zero relative separation ($r_{eh} = 0$), $\mathbf{k}$ is the reciprocal space wave-vector, $E_g$ denotes the bandgap of $MoS_2$, and $V(\mathbf{Q})$ is the electron-hole coulomb interaction. The dielectric screening modulates the following factors: (a) $|\psi(r_{eh} = 0)|^2 -$ due to a reduction in the 2D exciton binding energy[24,28]; (b) $E_g -$ due to bandgap renormalization effect in monolayer TMDs[29,30]; and (c) $V(\mathbf{Q}) -$ due to suppressed electron-hole interaction[31]. In Figure 3e, we show the variation in $J_{\mathbf{Q}}^{LR}$ with $\mathbf{Q}$ within the light cone, and hence the screening induced suppression of the long-range exchange in our samples (see Supplementary Note 4).

Another consequence of screening is the enhancement of the exciton radiative lifetime due to a reduced electron-hole wavefunction overlap. Here, a longer lifetime is undesirable as it leads to a larger valley decoherence (Figure 1a). To estimate the exciton lifetime and its role in the valley decoherence, we carry out TRPL measurements on our samples (see Methods and Supplementary Figure 11). The TRPL values obtained in our stacks are as follows: $< 5$ ps in the HMH stack, $6 - 8$ ps in the GHMHG stack, and $< 5$ ps in the GMG stack. The uncertainty in the lifetime in the HMH and GMG stacks arises as it is smaller than the 10% limit of our Instrument Response Function (IRF) width[32]. Nonetheless, several reports supporting these numbers are already available in literature[18,19,22,33]. Moreover, the qualitative trend in the exciton lifetime testifies the anticipated trend in Figure 1d, showing a clear enhancement in the exciton lifetime in the GHMHG stack, as compared to the other two stacks.

Obtaining an average DOLP of only 44.5 ($\pm 10$) % implies an ultra-short valley coherence time in the HMH stack, in agreement with previous reports[12,15]. On the other hand, in the GHMHG stack, an increased exciton lifetime is an evidence of screening induced enhancement of exciton lifetime as a result of introducing top and bottom FLG. Due to the opposing roles of reduced exchange and increased lifetime, we do not observe any improvement in the exciton DOLP in this sample compared with the HMH sample.

This side-effect of screening driven enhanced exciton lifetime is eliminated in the GMG stack through filtering, where light collection from the long-lived excitons is prohibited due to a fast transfer of excitons to graphene. The extracted timescale corresponds to the graphene-transfer-limited exciton lifetime in this system. This timescale is similar to that in the HMH stack, and in agreement with previous report[22,33]. Therefore, the significant exciton DOLP difference between the HMH and the GMG stacks is thus attributed to screening modified exchange interaction without any confounding effect due to a change in the exciton lifetime.

To obtain a quantitative understanding, we solve the steady-state form of the MSS equation:

$$\mathbf{G} = \frac{1}{\tau}\mathbf{S}(\mathbf{Q}) - \boldsymbol{\Omega}(\mathbf{Q}) \times \mathbf{S}(\mathbf{Q}) - \sum_{\mathbf{Q}'} \frac{w}{Q^2 \sin^2\frac{\alpha}{2}}[\mathbf{S}(\mathbf{Q}') - \mathbf{S}(\mathbf{Q})] \qquad (3)$$

We obtain the DOLP ($\langle S_x \rangle$) for the $A_{1s}$ exciton in the HMH and the GMG stack (Supplementary Note 4). For $W_{\mathbf{QQ}'}$, the exciton-impurity scattering rate expression is used (Supplementary Note 4.2). $\tau$ is the exciton lifetime, $\alpha$ is the angle between the initial ($\mathbf{Q}$) and the final ($\mathbf{Q}'$) state and $w$ is an overall scaling factor. We neglect the contribution of exciton-phonon scattering in decoherence at 5 K. Plotted in Figure 4 is the variation in $\langle S_x \rangle$ as a function of the $w$ for the two stacks. In both the cases, the V shaped variation is understood as follows: For small $w$, an increase in the scattering degrades the valley coherence due to enhanced exciton precession around $\boldsymbol{\Omega}$ (Figure 1a). However, this effect is non-monotonic, as on significantly enhancing the scattering rate, the DOLP starts increasing after reaching a minimum. This phenomenon is referred to as motional narrowing[10,34], and arises due to a cancellation of the accumulated randomness in the phase information of $\mathbf{S}$. Mathematically, the system is in the motional narrowing regime when the exciton scattering frequency becomes larger than the precession frequency, leading to a longer pseudospin coherence time.

We take the extracted homogeneous (Lorentzian component of the Voigt fitting) linewidth ($\Gamma_{hom}$) of the co-polarized exciton PL peak as the experimental analogue of the scattering rate. The experimental DOLP as a function of the Lorentzian linewidth, superimposed on the simulation results, is shown in Figure 4. Both in GMG and HMH stacks, the extracted value of $\Gamma_{hom}$ is much larger than the exciton lifetime limited linewidth, as obtained from TRPL. This suggests that the impurity scattering rate is similar in both the samples, and it dominates over other linewidth broadening mechanisms. This is also evident from the upward trend of the experimental DOLP with $\Gamma_{hom}$, which is in excellent agreement with the rising side of $\langle S_x \rangle$ versus $w$ in the simulation. This suggests that the whole operation lies in the motional narrowing regime in both the samples.

We conclude that the combined effect of enhanced screening, reduced lifetime due to interlayer transfer, and motional narrowing helps us to achieve $\sim 100\%$ exciton DOLP in our FLG-capped $MoS_2$ sample. This is a direct evidence of valley coherence being maintained during the entire exciton lifetime by cutting down the decoherence channels before the spontaneous emission. Achieving ~100% steady-state DOLP also indicates that the true valley coherence timescale beyond few picoseconds in the GMG stack, which is in excellent agreement with simulation result obtained by solving time-dependent MSS equation (see Supplementary Figure 12).

The combination of such ~100% polarization, coupled with background-free, narrow linewidth emission, makes the GMG stack a promising substrate for spectral diffusion-free, indistinguishable single photon source. As the initialized coherence in the exciton is shown to be staying protected for a longer time, the results have intriguing prospects on performing experiments involving coherent manipulation of exciton and building quantum system operating at these timescales.

## Methods

**Sample preparation:** All the stacks in this paper are prepared first by mechanically exfoliating the layered material on a Polydimethylsiloxane (PDMS) film, and then its subsequent transfer in a controlled manner underneath a microscope on a Si substrate covered with 285 nm thick thermally grown $SiO_2$. The thickness of the few-layer graphene is chosen to be 2-3 nm in all the stacks. The thickness of hBN layers is in the range of 20-30 nm apart from the GHMHG stack. To ensure strong screening in the GHMHG stack, the hBN thickness is kept at ~5 nm. After the preparation of the entire stack, the samples are annealed at $200^0$C for 5 hours (pressure ~$10^{-6}$ torr) to ensure better adhesion between successive layers.

**Sample characterization:** All the measurements are taken in a closed-cycle optical cryostat (Montana Instruments) at 5 K using a $\times 50$ long-working-distance objective having a numerical aperture of 0.5. To measure the exciton DOLP $\left[= \left(I_{H/H} - I_{\mathrm{H/V}}\right)/\left(I_{H/H} + I_{\mathrm{H/V}}\right)\right]$, we place the analyzer in the parallel $(I_{H/H})$ and perpendicular $(I_{\mathrm{H/V}})$ direction in the collection path relative to the excitation polarization direction. For the DOCP measurements, a quarter-wave plate is inserted just before the objective lens, and aligned at $45^0$ with respect to the incoming linearly polarized light. The time-resolved photoluminescence measurement is carried out using a 531 nm laser controlled by the PDL-800 D driver (laser pulse width is 48 ps). We use a single photon counting detector from Micro Photon Devices and the time correlated measurements are taken using the PicoHarp 300 TCSPC system (PicoQuant). We use a combination of two bandpass filters to get the time resolved counts of the $A_{1s}$ exciton at 5 K: 650 (FWHM - 55 nm) and 635 nm bandpass filter (FWHM - 10 nm) for the HMH and the GHMHG stack, and 650 (FWHM - 55 nm) and 660 nm bandpass filter (FWHM - 10 nm) for the GMG stack. The instrument response function (IRF) has an FWHM of 52 ps, and shows a decay of 23 ps. The deconvolution of the TRPL data with the IRF is carried out using the QuCoa software (PicoQuant).

**Acknowledgement**

This work was supported in part by a Core Research Grant from the Science and Engineering Research Board (SERB) under Department of Science and Technology (DST), a grant from Indian Space Research Organization (ISRO), a grant from MHRD under STARS, and a grant from MHRD, MeitY and DST Nano Mission through NNetRA. K.W. and T.T. acknowledge support from the JSPS KAKENHI (Grant Numbers 19H05790 and 20H00354).

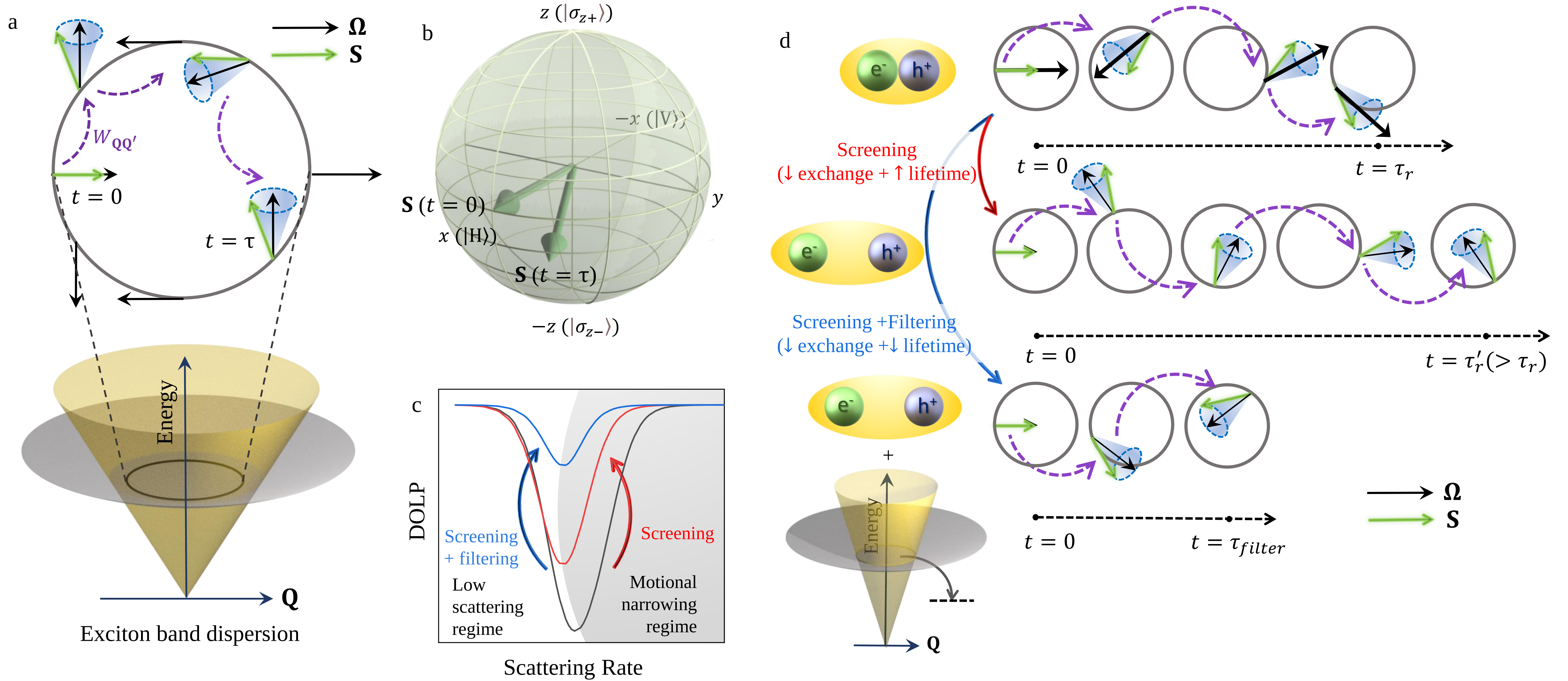

a
Ω
S
$W_{QQ'}$
$t = 0$
$t = \tau$
Energy
Q
Exciton band dispersion
b
$z\ (|\sigma_{z+}\rangle)$
$-x\ (|V\rangle)$
$\mathbf{S}\ (t = 0)$
$y$
$x\ (|H\rangle)$
$\mathbf{S}\ (t = \tau)$
$-z\ (|\sigma_{z-}\rangle)$
c
DOLP
Screening + filtering
Screening
Low scattering regime
Motional narrowing regime
Scattering Rate
d
Screening
(↓ exchange + ↑ lifetime)
Screening +Filtering
(↓ exchange +↓ lifetime)
$t = 0$
$t = \tau_r$
$t = \tau_r'(> \tau_r)$
$t = \tau_{filter}$
Ω
S

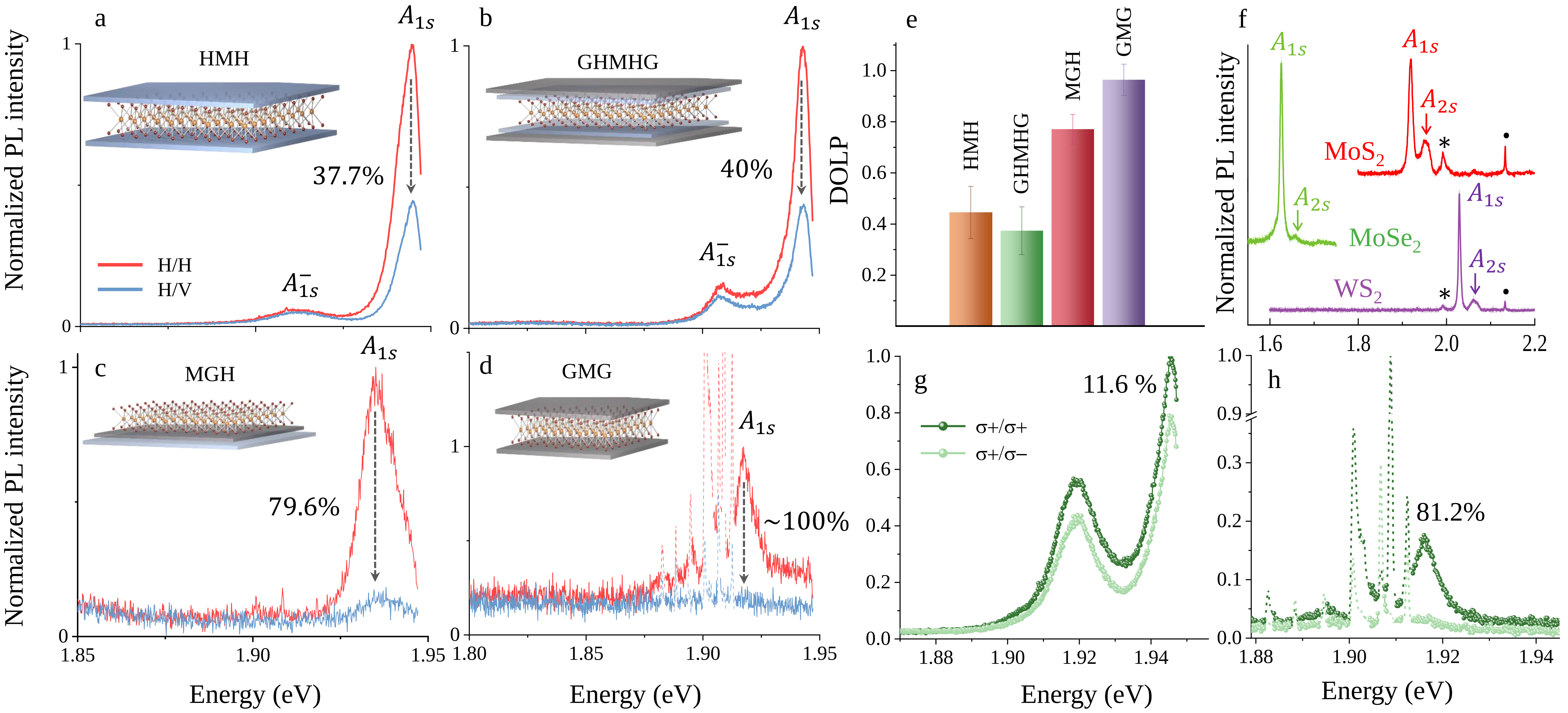

a
HMH
H/H
H/V
37.7%
$A_{1s}$
$A_{1s}^{-}$
b
GHMHG
40%
c
MGH
79.6%
d
GMG
~100%
e
DOLP
HMH
GHMHG
MGH
GMG
f
MoS$_2$
MoSe$_2$
WS$_2$
$A_{2s}$
g
σ+/σ+
σ+/σ−
11.6 %
h
81.2%
Normalized PL intensity
Energy (eV)

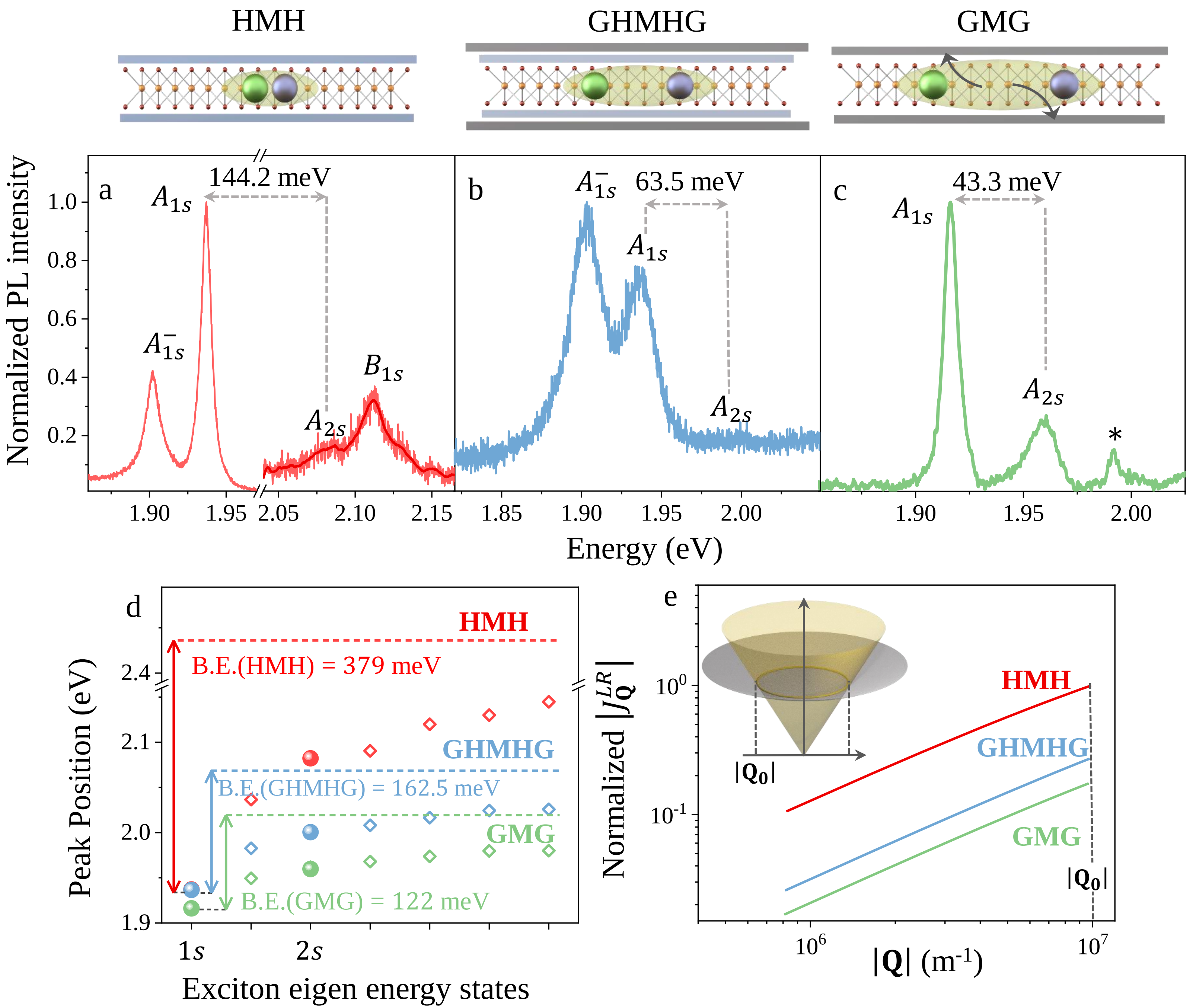

HMH
GHMHG
GMG
a
b
c
Normalized PL intensity
Energy (eV)
144.2 meV
63.5 meV
43.3 meV
$A_{1s}$
$A_{1s}^{-}$
$A_{2s}$
$B_{1s}$
*
d
HMH
B.E.(HMH) = 379 meV
GHMHG
B.E.(GHMHG) = 162.5 meV
GMG
B.E.(GMG) = 122 meV
Peak Position (eV)
1s
2s
Exciton eigen energy states
e
Normalized $|J_{\mathbf{Q}}^{LR}|$
$|\mathbf{Q_0}|$
$|\mathbf{Q}|$ ($m^{-1}$)
HMH
GHMHG
GMG

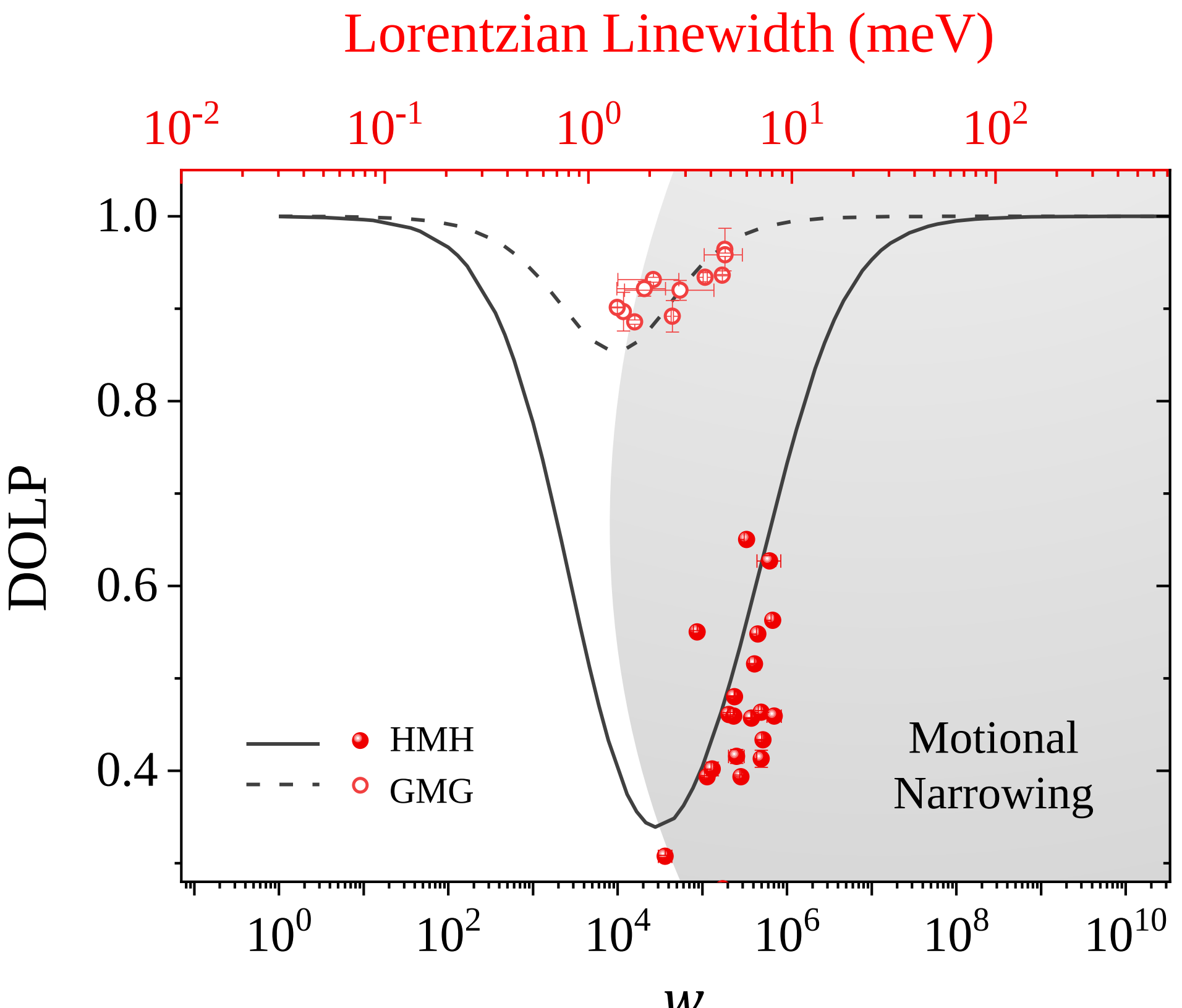

Lorentzian Linewidth (meV)
10-2
10-1
100
101
102
DOLP
1.0
0.8
0.6
0.4
HMH
GMG
Motional
Narrowing
100
102
104
106
108
1010
w

Figure 1: **Mechanism of exciton valley decoherence and the factors affecting it**. (a) Top view of a ring inside the light cone of the exciton band showing the exciton decoherence dynamics due to scattering $W_{\mathbf{QQ}'}$ within the light cone (purple dashed arrows) and subsequent precession because of inter-valley exchange induced pseudo-magnetic field (black solid arrows). (b) The direction of $\mathbf{S}$ at generation ($t = 0$) and recombination ($t = \tau$) are shown in the Bloch sphere. (c) The calculated DOLP as a function of scattering rate with motional narrowing regime shown in shade. (d) The decoherence of $\mathbf{S}$ (green arrows) with time due to scattering (dashed purple arrow) and precession around $\mathbf{\Omega}$ (black arrows) for an exciton for three different scenarios. The effect of screening versus screening + filtering on valley decoherence is compared in the three rows.

Figure 2: **Excitonic emission from different stacks along with DOLP and DOCP.** (a-d) PL spectra with near-resonant 633 nm linearly polarized excitation in co-(H/H) and cross- (H/V) polarized detection configuration in the (a) hBN-$MoS_2$-hBN (HMH) stack, (b) FLG-hBN-$MoS_2$-hBN-FLG (GHMHG) stack, (c) $MoS_2$-FLG-hBN (MGH) stack, and the (d) FLG-$MoS_2$-FLG (GMG) stack at $T = 5$ K. $A_{1s}$ and $A_{1s}^{-}$ represent the exciton and the trion peak, respectively. (e) Bar graph comparing the experimental DOLP values along with the error bars in the four stacks. (f) PL spectra obtained from FLG-TMD-FLG stack (using 532 nm excitation) for monolayer $MoS_2$, $MoSe_2$, and $WS_2$ showing the prominent $A_{1s}$ and $A_{2s}$ peaks. The $A_{2s} - A_{1s}$ separation is around 44 (32) meV in $MoS_2$ ($MoSe_2$, $WS_2$). The peaks marked as $*$ and $\cdot$ are the 2D and the G Raman peaks of the FLG. (g-h) Representative PL spectra taken with circularly polarized excitation in co- ($\sigma+/\sigma+$) and cross- ($\sigma+/\sigma-$) polarized detection configuration. The corresponding DOCP value of the $A_{1s}$ exciton in (g) the HMH and (h) the GMG stack at $T = 5$ K is shown in the inset. The peaks indicated by the dashed lines in (d) and (h) represent the prominent Raman peaks in the GMG stack due to dual resonance.

Figure 3: **Evidence of graphene induced screening of electron-hole exchange interaction.** (a-c) Top panel: Schematic representation of the screening effect in the different stacks. Bottom panel: PL spectra taken with 532 nm excitation highlighting the different degree of dielectric screening in our samples. The $A_{2s} - A_{1s}$ separation for the HMH stack (144.5 meV), GHMHG stack (63.5 meV), and the GMG stack (43.3 meV) is indicated by the dashed arrows. $*$ in (c) is the 2D Raman peak of FLG. (d) Eigen energies (open circle) and the corresponding binding energies of the $A_{1s}$ exciton obtained from the Bethe-Salpeter equation. The solid symbols

denote experimental data. The dashed lines are the corresponding continuum levels. (e) Calculated value of the normalized long-range exchange potential variation inside the light cone ($|\mathbf{Q}| < |\mathbf{Q_0}|$). Inset: light emitting region ($|\mathbf{Q}| < |\mathbf{Q_0}|$) of the exciton band highlighted by the light cone.

Figure 4: **Comparison of the experimental data with the steady-state solution of the MSS equation.** Simulation results comparing the exciton DOLP ($\langle S_x \rangle$) as a function of the scaling factor ($w$) of the scattering rate (bottom axis) for the HMH stack (solid black trace) and the GMG stack (dashed black trace). The downward trend in the left-hand side (unshaded region) is the low-scattering regime, and the upward trend in the right-hand side (shaded region) is the motional narrowing regime. Overlapped on the simulation results is the experimentally obtained DOLP variation with the Lorentzian linewidth ($\Gamma_{hom}$) of the corresponding co-polarized PL spectrum (top axis) for the HMH stack (solid spheres) and the GMG stack (open circles). The upward trend of the experimental data suggests that both the stacks are operating in the motional narrowing regime.

# Supplementary Information: Observation of ~100% valley-coherent excitons in monolayer $MoS_2$ through giant enhancement of valley coherence time

Garima Gupta[1], Kenji Watanabe[2], Takashi Taniguchi[3], and Kausik Majumdar[1,*]

[1]*Department of Electrical Communication Engineering, Indian Institute of Science, Bangalore 560012, India*

[2] *Research Center for Functional Materials, National Institute for Materials Science, 1-1 Namiki, Tsukuba 305-044, Japan*

[3] *International Center for Materials Nanoarchitectonics, National Institute for Materials Science, 1-1 Namiki, Tsukuba 305-044, Japan*

**corresponding author, E-mail: kausikm@iisc.ac.in*

**Supplementary Note 1: Derivation of the Maialle-Silva-Sham (MSS) equation**

Consider an exciton pseudospin **S** pointing in the $(\theta, \phi)$ direction on the Bloch sphere as shown below.

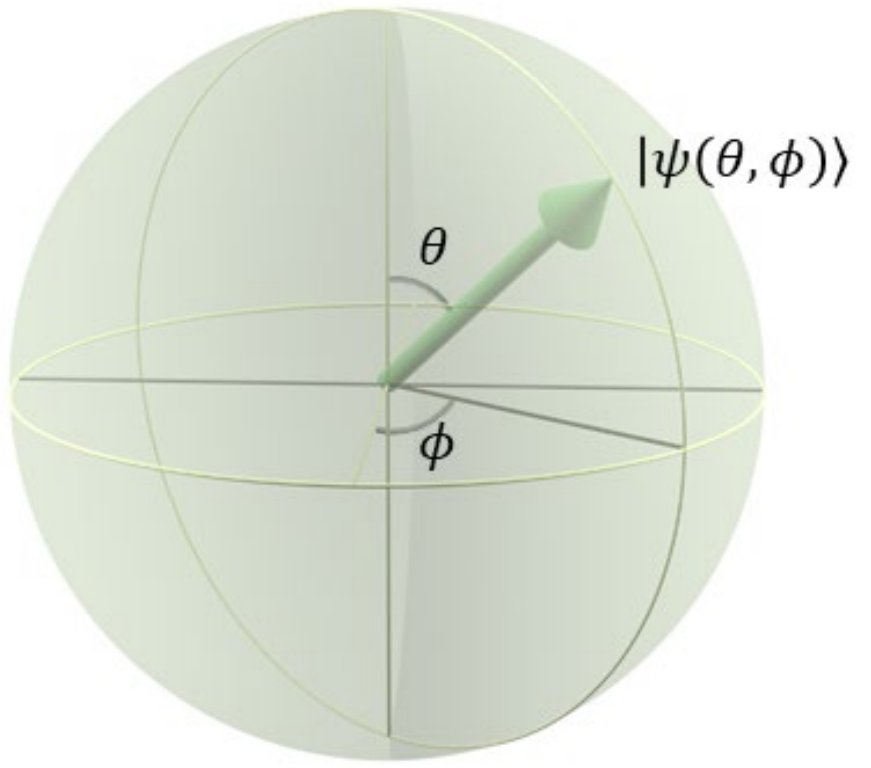

**Supplementary Figure 1:** A schematic representation of the Bloch sphere indicating the exciton pseudospin direction

This exciton state can be written in the basis of polar states $|\sigma_{z+}\rangle$ and $|\sigma_{z-}\rangle$ as:

$$|\psi(\theta,\phi)\rangle = \begin{pmatrix} \cos\frac{\theta}{2} \\ e^{i\phi}\sin\frac{\theta}{2} \end{pmatrix}$$

The density matrix form of this pure state single qubit system is given by:

$$\rho(\theta,\phi) = |\psi(\theta,\phi)\rangle\langle\psi(\theta,\phi)|$$

which may be written in terms of the identity matrix I and Pauli Matrices as:

$$\rho(\theta,\phi) = \frac{1}{2}\left(\mathrm{I} + \sin\theta\cos\phi\ \sigma_x + \sin\theta\sin\phi\ \sigma_y + \cos\theta\ \sigma_z\right)$$

We denote the projection of **S** along the $x$, $y$ and $z$ axis of the Bloch sphere as $S_x = \sin\theta\cos\phi$, $S_y = \sin\theta\sin\phi$ and $S_z = \cos\theta$. $S_x$, $S_y$ and $S_z$ are the coefficients representing the expectation value of angular momentum along that direction. In terms of polarization, the $x$, $y$ axes on the equator of the Bloch sphere correspond to the linearly polarized light along $x$ axis and $45^0$ from $x$ axis, and the polar points correspond to the circularly polarized light in the real space.

The time evolution of the density operator $\rho$ for the mixed state in the presence of exchange field and momentum scattering $\left(\mathrm{W}_{\mathbf{QQ}'}\right)$ $is$ given by MSS as[1]:

$$\frac{d\rho(\mathbf{Q},t)}{dt} = \frac{i}{\hbar}[\rho(\mathbf{Q},t),H] + \sum_{\mathbf{Q}'} \mathrm{W}_{\mathbf{QQ}'}\,[\rho(\mathbf{Q}',t) - \rho(\mathbf{Q},t)] - \frac{\rho(\mathbf{Q},t)}{\tau} + G \qquad (1)$$

Here $\tau$ represents the net exciton lifetime within the light cone, and the matrix $G$ accounts for the exciton generation rate. $H$ is the total pseudospin Hamiltonian which can be decomposed into $H = H_0 + H_1(\mathbf{Q})$, where $H_0$ is the diagonal matrix that accounts for the applied magnetic field in the out-of-plane direction:

$$H_0 = \begin{bmatrix} E_+ & 0 \\ 0 & E_- \end{bmatrix}$$

and $H_1$ is the matrix containing the off-diagonal elements that is responsible for the exchange driven exciton pseudospin flip:

$$H_1(\mathbf{Q}) = \frac{\hbar\Omega_{\parallel}(\mathbf{Q})}{2} \begin{bmatrix} 0 & e^{-i2\phi} \\ e^{i2\phi} & 0 \end{bmatrix}$$

where $\phi$ is the angle between $\mathbf{Q}$ and the $x$ axis.

$H_0$ can be expanded as

$$H_0 = \frac{1}{2}Tr(H_0)\mathrm{I} + \frac{\hbar\Omega_0}{2}\sigma_z$$

where $\Omega_0 = (E_+ - E_-)/\hbar$, and $H_1(\mathbf{Q})$ can be written as

$$H_1(\mathbf{Q}) = \frac{\hbar\,\Omega_{\parallel}(\mathbf{Q})}{2}\left[\cos(2\phi)\,\sigma_x + \sin(2\phi)\,\sigma_y\right].$$

Thus, we obtain

$$H = \frac{1}{2}Tr(H_0)\mathrm{I} + \frac{\hbar}{2}\boldsymbol{\Omega}.\boldsymbol{\sigma}$$

where $\boldsymbol{\Omega} = [\Omega_{\parallel}\cos\phi,\ \Omega_{\parallel}\sin\phi,\ \Omega_0]$. Similarly, decomposing the density matrix in terms of its trace $N$ and the traceless part $\mathbf{S}.\boldsymbol{\sigma}$, where $\boldsymbol{\sigma}$ denotes the Pauli matrices, the commutator $[\rho(\mathbf{Q},t),H\,]$ turns out to be:

$$[\rho(\mathbf{Q},t),H] = \frac{\hbar}{2}[\mathbf{S}(\mathbf{Q},t).\boldsymbol{\sigma},\boldsymbol{\Omega}.\boldsymbol{\sigma}] \qquad (2)$$

Combining equations (1) and (2):

$$\frac{d}{dt}\left(\frac{N(\mathbf{Q},t)}{2}\mathrm{I}+\mathbf{S}(\mathbf{Q},t).\boldsymbol{\sigma}\right) = \frac{i}{2}[\mathbf{S}(\mathbf{Q},t).\boldsymbol{\sigma},\boldsymbol{\Omega}.\boldsymbol{\sigma}]$$

$$+\sum_{\mathbf{Q}'}\mathrm{W}_{\mathbf{QQ}'}\left(\frac{N(\mathbf{Q}',t)-N(\mathbf{Q},t)}{2}\mathrm{I}+\ \mathbf{S}(\mathbf{Q}',t).\boldsymbol{\sigma}-\mathbf{S}(\mathbf{Q},t).\boldsymbol{\sigma}\right)$$

$$-\frac{1}{\tau}\left(\frac{N(\mathbf{Q},t)}{2}\mathrm{I}+\mathbf{S}(\mathbf{Q},t).\boldsymbol{\sigma}\right)+G$$

On equating the traceless matrices on both the sides of the above equation, we get:

$$\frac{d}{dt}\left(\mathbf{S}(\mathbf{Q},t).\boldsymbol{\sigma}\right) = \frac{i}{2}[\mathbf{S}(\mathbf{Q},t).\boldsymbol{\sigma},\boldsymbol{\Omega}.\boldsymbol{\sigma}]$$

$$+\sum_{\mathbf{Q}'}\mathrm{W}_{\mathbf{QQ}'}\left(\mathbf{S}(\mathbf{Q}',t).\boldsymbol{\sigma}-\mathbf{S}(\mathbf{Q},t).\boldsymbol{\sigma}\right)-\frac{1}{\tau}\left(\mathbf{S}(\mathbf{Q},t).\boldsymbol{\sigma}\right)+\mathbf{G}.\boldsymbol{\sigma}$$

where $\mathbf{G}=[G_x\ G_y\ G_z]$ is the generation rate vector.

By expanding the commutation bracket and using the vector identity $(\mathbf{S}.\boldsymbol{\sigma})(\boldsymbol{\Omega}.\boldsymbol{\sigma}) = (\mathbf{S}.\boldsymbol{\Omega})\mathrm{I}+i(\mathbf{S}\times\boldsymbol{\Omega}).\boldsymbol{\sigma}$, we get the final form of the MSS equation as:

$$\frac{d\mathbf{S}(\mathbf{Q})}{dt}=\boldsymbol{\Omega}(\mathbf{Q})\times\mathbf{S}(\mathbf{Q})+\sum_{\mathbf{Q}'}\mathrm{W}_{\mathbf{QQ}'}\left[\mathbf{S}(\mathbf{Q}')-\mathbf{S}(\mathbf{Q})\right]-\frac{1}{\tau}\mathbf{S}(\mathbf{Q})+\mathbf{G}$$

**Supplementary Note 2: Derivation that DOLP is given by $\langle S_x\rangle$**

The state $|\psi\rangle$ in the basis states of two-orthogonal circular polarizations $|\sigma_{z+}\rangle, |\sigma_{z-}\rangle$ is expressed as:

$$|\psi\rangle=\cos\frac{\theta}{2}|\sigma_{z+}\rangle+\sin\frac{\theta}{2}e^{i\phi}|\sigma_{z-}\rangle$$

Thus, the degree of circular polarization (DOCP) of this single exciton qubit state can be obtained as:

$$\mathrm{DOCP}=\frac{\left(\cos^2\frac{\theta}{2}-\sin^2\frac{\theta}{2}\right)}{\left(\cos^2\frac{\theta}{2}+\sin^2\frac{\theta}{2}\right)}=\cos\theta=S_z$$

Similarly, in the basis states of the $|\sigma_{x+}\rangle, |\sigma_{x-}\rangle$ vectors that represent the two orthogonal linear polarization directions [vertical (V) and horizontal (H)], the state can also be represented as

$$|\psi\rangle = \frac{1}{\sqrt{2}}\left(\cos\frac{\theta}{2} + \sin\frac{\theta}{2}e^{i\phi}\right)|\sigma_{x+}\rangle + \frac{1}{\sqrt{2}}\left(\cos\frac{\theta}{2} - \sin\frac{\theta}{2}e^{i\phi}\right)|\sigma_{x-}\rangle$$

in which case, the degree of linear polarization (DOLP) can be calculated as:

$$\text{DOLP} = \frac{\left(\left|\cos\frac{\theta}{2} + \sin\frac{\theta}{2}e^{i\phi}\right|^2 - \left|\cos\frac{\theta}{2} - \sin\frac{\theta}{2}e^{i\phi}\right|^2\right)}{\left(\left|\cos\frac{\theta}{2} + \sin\frac{\theta}{2}e^{i\phi}\right|^2 + \left|\cos\frac{\theta}{2} - \sin\frac{\theta}{2}e^{i\phi}\right|^2\right)} = \sin\theta\cos\phi = S_x$$

The overall steady state DOLP is then obtained as $\langle S_x\rangle$ averaged over the **Q** space.

**Supplementary Note 3: DOLP is greater than DOCP in a 2D system**

To show that the exciton DOLP is generally higher than its DOCP value in 2D semiconductors, we take a special case of no scattering scenario ($\mathrm{W}_{\mathbf{QQ'}} \to 0$). The steady-state form of the MSS equation is then given by:

$$\mathbf{G} = \frac{1}{\tau}\mathbf{S}(\mathbf{Q}) - \mathbf{\Omega}(\mathbf{Q}) \times \mathbf{S}(\mathbf{Q})$$

$\mathbf{\Omega} = [\Omega_{\parallel}\cos\phi,\ \Omega_{\parallel}\sin\phi, 0]$ in the absence of an external magnetic field, and $\phi$ is the azimuthal angle between **Q** and $x$ axis.

The steady state value of $S_z$ can be obtained by putting $G_x = 0, G_y = 0$ *and* $G_z = G$ in the above equation as

$$S_z(\mathbf{Q}) = \frac{G\tau}{(1 + \tau^2\Omega_{\parallel}^2(\mathbf{Q}))}$$

Similarly, the steady state value of $S_x$ can be obtained by putting $G_x = G, G_y = 0$ *and* $G_z = 0$:

$$S_x(\mathbf{Q}) = \frac{G\tau\,(1 + \tau^2\Omega_{\parallel}^2(\mathbf{Q})\cos^2 2\phi)}{(1 + \tau^2\Omega_{\parallel}^2(\mathbf{Q}))}$$

After the initial generation along the $0^0, 180^0$ axis, the degree of linear polarization $S_x(\mathbf{Q})$ at those **Q** values in the limiting condition of $\mathrm{W}_{\mathbf{QQ'}} \to 0$ ($\phi \to 0^0, 180^0$) is $G\tau$. Hence, we can see that, in the case of no scattering, the pseudospin $S_x$ is nearly independent of $\Omega_{\parallel}$, whereas $S_z$ suffers due to $\Omega_{\parallel}$. A physical explanation for this is that the linear polarization degrades because of one of the in-plane magnetic field components, but the out-of-plane circular polarization experiences the effect of both the in-plane components of the exchange magnetic field[1,2]. As a result, the DOLP is generally higher than the DOCP in these 2D exciton.

**Supplementary Note 4: Simulation Details**

**4.1. Electronic and Excitonic band structure calculation**

We use $2 \times 2$ **k.p** Hamiltonian to get the band dispersion of the lowest energy conduction band (CB) and highest energy valence band (VB) in the $\mathbf{K}, \mathbf{K}'$ valley in monolayer TMDs. The Hamiltonian in the atomic orbital basis states of $|d_{z^2}\rangle$ and $\frac{1}{\sqrt{2}}\left(|d_{x^2-y^2}\rangle + i\tau|d_{xy}\rangle\right)$, where $\tau$ ($= \pm 1$) represents the valley index, is given by[3]:

$$\begin{bmatrix} E_g & at(\tau k_x - ik_y) \\ at(\tau k_x + ik_y) & 0 \end{bmatrix}$$

$k_x, k_y$ are the wave vectors in the reciprocal space. The value of the lattice constant $a$, the hopping amplitude $t$, and the quasi-particle band gap $E_g$ for calculating monolayer $MoS_2$ band structure are taken as 3.193 Å, 1.10 eV and 2.4 eV, respectively. The eigen energies (eigen functions) obtained after solving the above Hamiltonian is denoted as $E_{c,\mathbf{k}}(|c,\mathbf{k}\rangle)$, $E_{v,\mathbf{k}}(|v,\mathbf{k}\rangle)$ for the conduction and valence band, respectively.

In the basis of the combined CB ($|c, \mathbf{k}+\mathbf{Q}\rangle$) and VB ($|v,\mathbf{k}\rangle$) pair states, denoted together as $|vc\mathbf{kQ}\rangle$, where $\mathbf{Q}$ ($= \mathbf{k_e} + \mathbf{k_h}$) is the exciton center-of-mass momentum, the exciton band structure is calculated by solving the following Bethe-Salpeter equation[4]:

$$\langle vc\mathbf{kQ}|H|vc\mathbf{k}'\mathbf{Q}\rangle = \delta_{kk'}\left(E_{c,\mathbf{k+Q}} - E_{v,\mathbf{k}}\right) - (D - X)(\mathbf{k}, \mathbf{k}', \mathbf{Q})$$

$D, X$ are the direct and the exchange interaction terms. We neglect the contribution of exchange interaction at small $\mathbf{Q}$ within the light cone in the exciton band structure calculation. The direct coulombic interaction is given by:

$$D = \frac{1}{A} V_{k-k'} \langle c, \mathbf{k}+\mathbf{Q}|c, \mathbf{k}'+\mathbf{Q}\rangle \langle v, \mathbf{k}'|v, \mathbf{k}\rangle$$

Here $A$ is the area of the two-dimensional crystal and $V_{k-k'}$ is the interaction potential given by[5]:

$$V_q = \frac{2\pi {q_0}^2 e^{-q\xi}}{q} \frac{1}{\epsilon(q)}$$

Here $q_0$ is the magnitude of the charge of an electron, $q = |\mathbf{k} - \mathbf{k}'|$, and $\xi$ is a fitting parameter that reflects the extension of the wavefunction of the electron and the hole in the out-of-plane direction. The dielectric function $\epsilon(q)$ is given by

$$\epsilon(q) = \frac{(1 - p_b p_t e^{-2\eta q D})\kappa}{(1 - p_t e^{-\eta q D})(1 - p_b e^{-\eta q D})} + r_0 q e^{-q\xi}$$

For the case of identical dielectric environment on top and bottom, $p_b = p_t = P = (\epsilon_{ENV} - \kappa)/(\epsilon_{ENV} + \kappa)$. $D$ is the thickness of the monolayer sheet, $\eta = \sqrt{\epsilon_{\parallel}/\epsilon_{\perp}}$ and $\kappa = \sqrt{\epsilon_{\parallel}\epsilon_{\perp}}$. $\epsilon_{\parallel}(\epsilon_{\perp})$ and $\epsilon_{ENV}$ is the effective in-plane (out-of-plane) and environmental dielectric constant of the monolayer TMD. We use the high frequency dielectric constant value of $\epsilon_{\parallel} = 5.01$ and $\epsilon_{\perp} = 6.07$ in our calculations.

The fitting parameters for the HMH, GHMHG and the GMG stacks are summarized below:

| | $\epsilon_{ENV}$ | $r_0$ (m) | $\xi$(Å) |
|---|---|---|---|
| HMH | 4 | $12 \times 10^{-8}$ | 8.64 |
| GHMHG | 16 | $23.6 \times 10^{-8}$ | 20 |
| GMG | 20 | $32 \times 10^{-8}$ | 25.6 |

With these fitting parameters, we ensure that the calculated $A_{2s} - A_{1s}$ energy separation matches with the experimentally obtained value. The exciton energy spectrum for the first few states showing the $A_{1s}$ exciton binding energy change in the three stacks is plotted in the main text (Figure 3d), and the full spectrum is shown below.

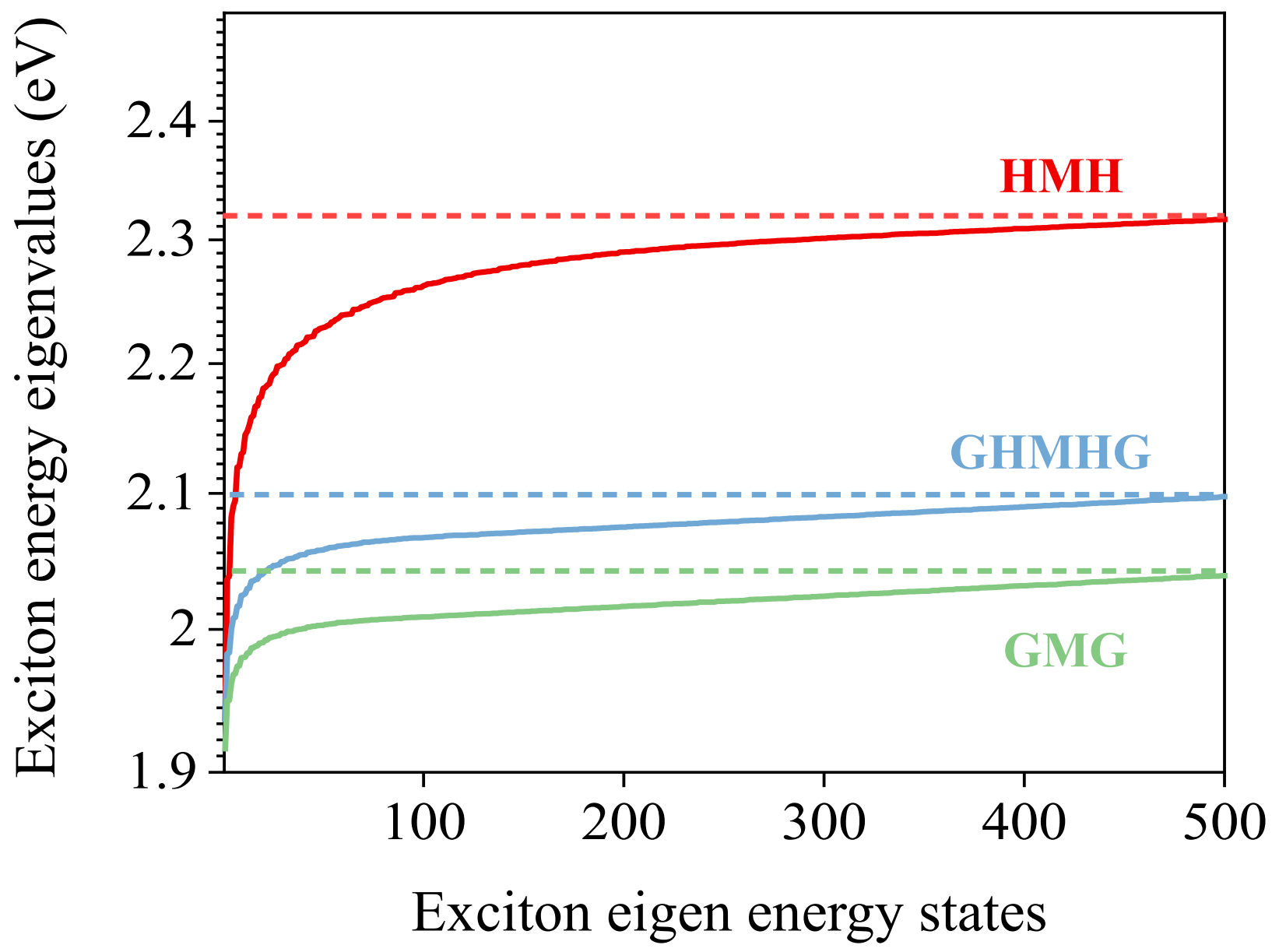

**Supplementary Figure 2:** The calculated exciton eigen energy spectrum (solid lines) for the HMH (in red), GHMHG (in blue), and the GMG (in green) stack obtained after solving the Bethe-Salpeter equation. The dashed lines represent the respective continuum energy levels. The extracted $A_{1s}$ exciton binding energy is 379 meV in the HMH stack. The binding energy reduces to 162.5 and 122 meV in the GHMHG and the GMG stack, respectively due to dielectric screening.

### 4.2. DOLP calculation

The DOLP for the three stacks is calculated using the steady-state form of the MSS equation given as:

$$\mathbf{G} = \frac{1}{\tau}\mathbf{S}(\mathbf{Q}) - \boldsymbol{\Omega}(\mathbf{Q}) \times \mathbf{S}(\mathbf{Q}) - \sum_{\mathbf{Q}'} W_{\mathbf{QQ}'}[\mathbf{S}(\mathbf{Q}') - \mathbf{S}(\mathbf{Q})]$$

$\tau$ is the net exciton lifetime, $\mathbf{G}$ is the exciton generation rate, $W_{\mathbf{QQ}'}$ is the exciton scattering rate within the light cone, and $\Omega_{\parallel}(\mathbf{Q})$ $(= 2|J_{\mathbf{Q}}^{LR}|/\hbar)$ is the precession frequency magnitude due to intervalley exchange induced magnetic field, where $J_{\mathbf{Q}}^{LR}$ is the long-range component of exchange, and is given by[6]:

$$J_{\mathbf{Q}}^{LR} = -\frac{|\sum_{\mathbf{k}} \psi(\mathbf{k})|^2}{A} \frac{a^2 t^2}{E_g^2} V(\mathbf{Q}) \, |\mathbf{Q}|^2 e^{-2i\phi}$$

$\phi$ is the azimuthal angle in the $\mathbf{Q}$-space. The screening modifies the exchange interaction through the change in the electron-hole wavefunction overlap $|\sum_{\mathbf{k}} \psi(\mathbf{k})|^2/A$, bandgap $E_g$ and the Fourier potential $V(\mathbf{Q})$. In Figure 3e of the main text, the magnitude of the exchange interaction potential is plotted as a function of $|\mathbf{Q}|$ within the light cone. During the calculation, we fit the parameters such that the $A_{2s} - A_{1s}$ separation remains the same as that obtained experimentally. $\tau$ is the exciton lifetime. For $\mathrm{W}_{\mathbf{QQ'}}$, we take the exciton-impurity scattering within the light cone as the only scattering mechanism at 5 K, neglecting exciton-phonon scattering. For simplicity, to get the exciton-impurity scattering matrix element, we use a similar form of the perturbing potential expression given for electron-ionized impurity scattering. We use the generic screened coulomb potential form as[7]

$$\frac{{q_0}^2}{4\pi\epsilon_0\epsilon_s r} \, e^{-r/L_D}$$

$\epsilon_0$ is the dielectric permittivity of vacuum and $L_D$is the Debye length. The matrix element for the above perturbing potential for an exciton scattering from $\mathbf{Q}$ to $\mathbf{Q'}$ state is given by

$$\begin{aligned} H_{\mathbf{Q'Q}} &= \frac{1}{A}\left(\frac{{q_0}^2}{4\pi\epsilon_0\epsilon_s r}\right) \int e^{-i\mathbf{Q'}.\mathbf{r}} \, \frac{e^{-r/L_D}}{r} \, e^{i\mathbf{Q}.\mathbf{r}} \, d^2\mathbf{r} \\ &= \left(\frac{{q_0}^2}{4\pi\epsilon_0\epsilon_s A}\right) \iint e^{i(\mathbf{Q}-\mathbf{Q'}).\mathbf{r}} \, e^{-r/L_D} \, dr \, d\phi \end{aligned}$$

Putting $\mathbf{Q'} - \mathbf{Q} = \boldsymbol{\beta}$ (note that, $|\mathbf{Q'}| = |\mathbf{Q}|$ since Coulomb scattering is elastic in nature), we obtain

$$H_{\mathbf{Q'Q}} = \left(\frac{{q_0}^2}{4\pi\epsilon_0\epsilon_s A}\right) \iint e^{-i\beta r\cos\phi} \, e^{-r/L_D} dr \, d\phi$$

The above expression after integration over $r$ results in

$$H_{\mathbf{Q'Q}} = \left(\frac{{q_0}^2 L_D}{4\pi\epsilon_0\epsilon_s A}\right)\left[\int \frac{1}{1+\beta^2 L_D^2 \cos^2\phi} \, d\phi \; - i \int \frac{\beta L_D \cos\phi}{1+\beta^2 L_D^2 \cos^2\phi} \, d\phi\right]$$

The result of the first integration is $2\pi/\sqrt{1+\beta^2 L_D^2}$ and the second one is zero. The matrix element thus becomes

$$H_{\mathbf{Q'Q}} = \frac{{q_0}^2}{2\epsilon_0\epsilon_s A\sqrt{\beta^2 + 1/L_D^2}}$$

We assume the limiting case of $L_D \rightarrow \infty$ and obtain the following expression for the scattering rate $\mathrm{W}_{\mathbf{Q'Q}}$ as

$$\mathrm{W}_{\mathbf{Q'Q}} \propto \left|H_{\mathbf{Q'Q}}\right|^2 = \left(\frac{{q_0}^2}{2\epsilon_0\epsilon_s A\,\beta}\right)^2$$

where $\beta = \frac{2Q}{\hbar}\sin\alpha/2$, $\alpha$ is the angle between the initial $(\mathbf{Q})$ and the final $(\mathbf{Q}')$ exciton state. The final form of the steady state MSS equation after plugging in the above form of scattering rate $\mathrm{W}_{\mathbf{Q'Q}}$ as given in the main text is given as:

$$\mathbf{G} = \frac{1}{\tau}\mathbf{S}(\mathbf{Q}) - \mathbf{\Omega}(\mathbf{Q}) \times \mathbf{S}(\mathbf{Q}) - \sum_{\mathbf{Q}'} \frac{w}{Q^2 \sin^2\frac{\alpha}{2}}[\mathbf{S}(\mathbf{Q}') - \mathbf{S}(\mathbf{Q})]$$

where $w$ is used as a fitting parameter. Finally, the DOLP $(= \langle S_x \rangle)$ is calculated by averaging the calculated pseudospin at the $\mathbf{Q}$-values within the light cone and is plotted in Figure 4 in the main text for the HMH and the GMG sample.

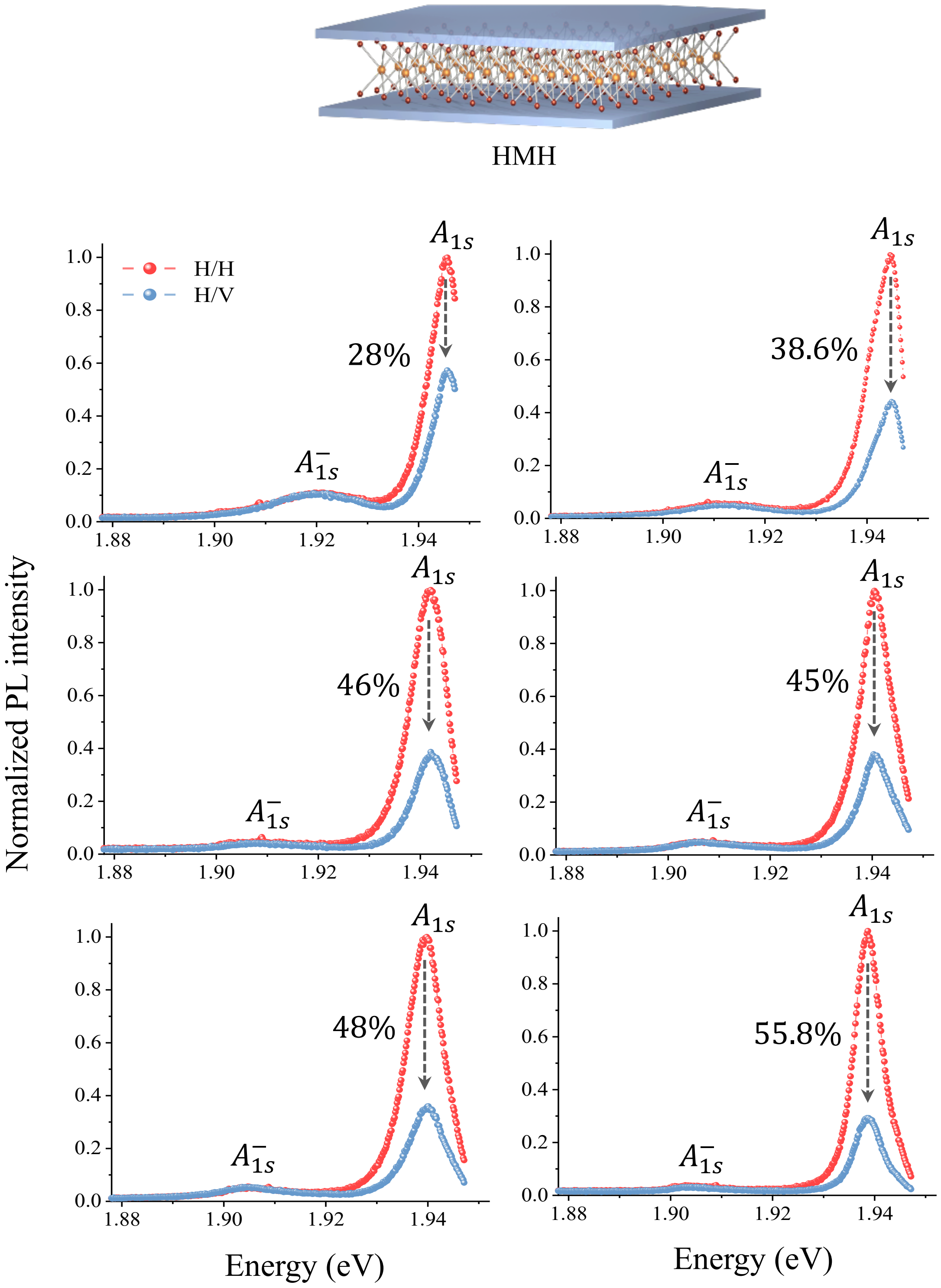

**Supplementary Figure 3:** The co- (H/H, in red) and cross- (H/V, in blue) linearly polarized PL spectra obtained at different spots on the HMH stack in the increasing order of DOLP. The range of the DOLP values obtained lies between $26 - 64$ %.

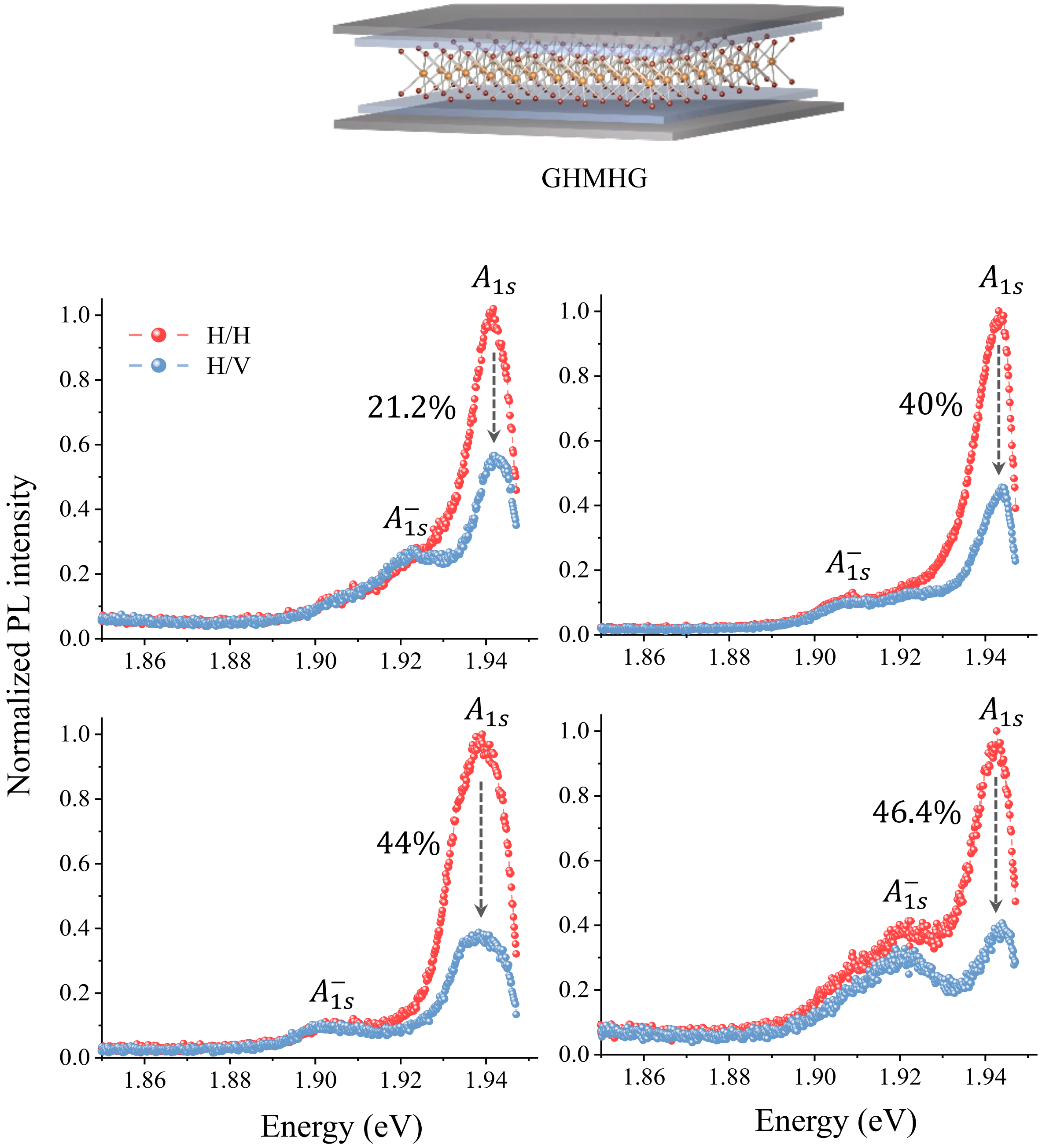

**Supplementary Figure 4:** The co- (H/H, in red) and cross- (H/V, in blue) linearly polarized PL spectra obtained at different spots on the GHMHG stack. The DOLP values range within $21 - 46$ % in this stack.

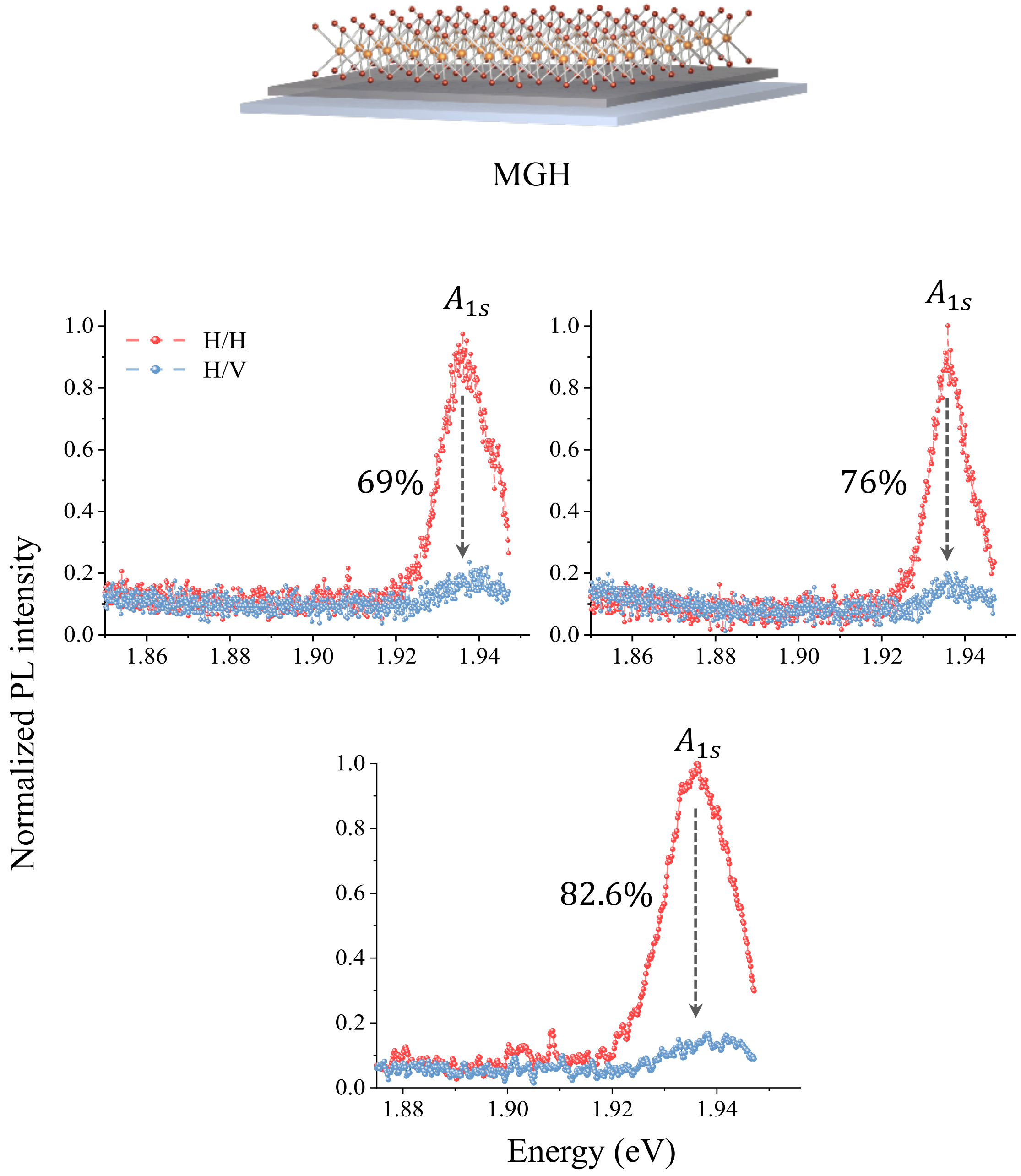

**Supplementary Figure 5:** The co- (H/H, in red) and cross- (H/V, in blue) linearly polarized PL spectra obtained at different spots on the MGH stack. The DOLP values range within 69 − 82 % in this stack.

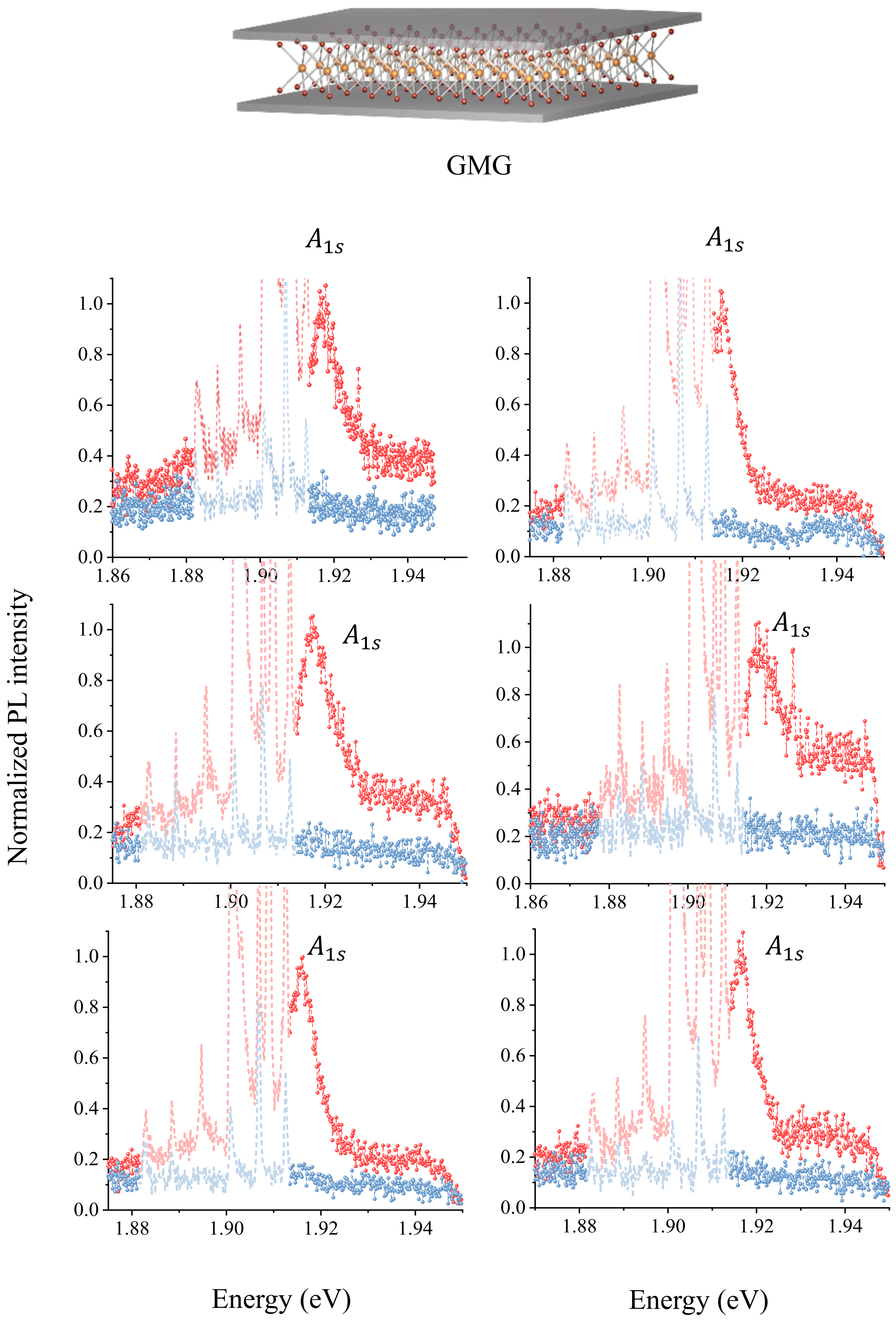
GMG
$A_{1s}$
Normalized PL intensity
Energy (eV)
Energy (eV)

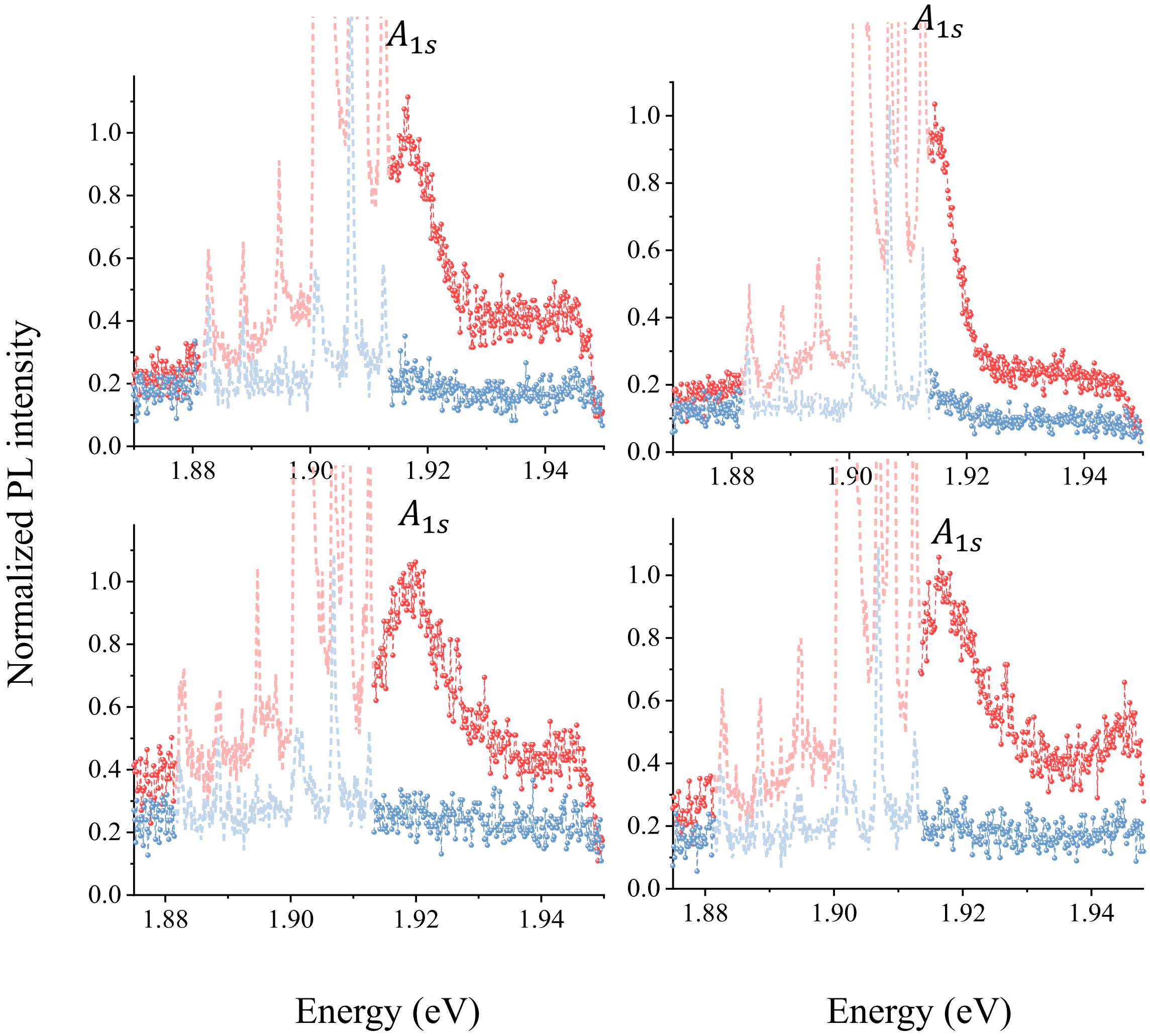

**Supplementary Figure 6:** The co- (H/H, in red) and cross- (H/V, in blue) linearly polarized PL spectra taken at different spots on the GMG showing 100% DOLP. The peaks indicated by the dashed lines represent Raman peaks. The dashed lines are the strong Raman peaks due to dual resonance in the GMG stack.

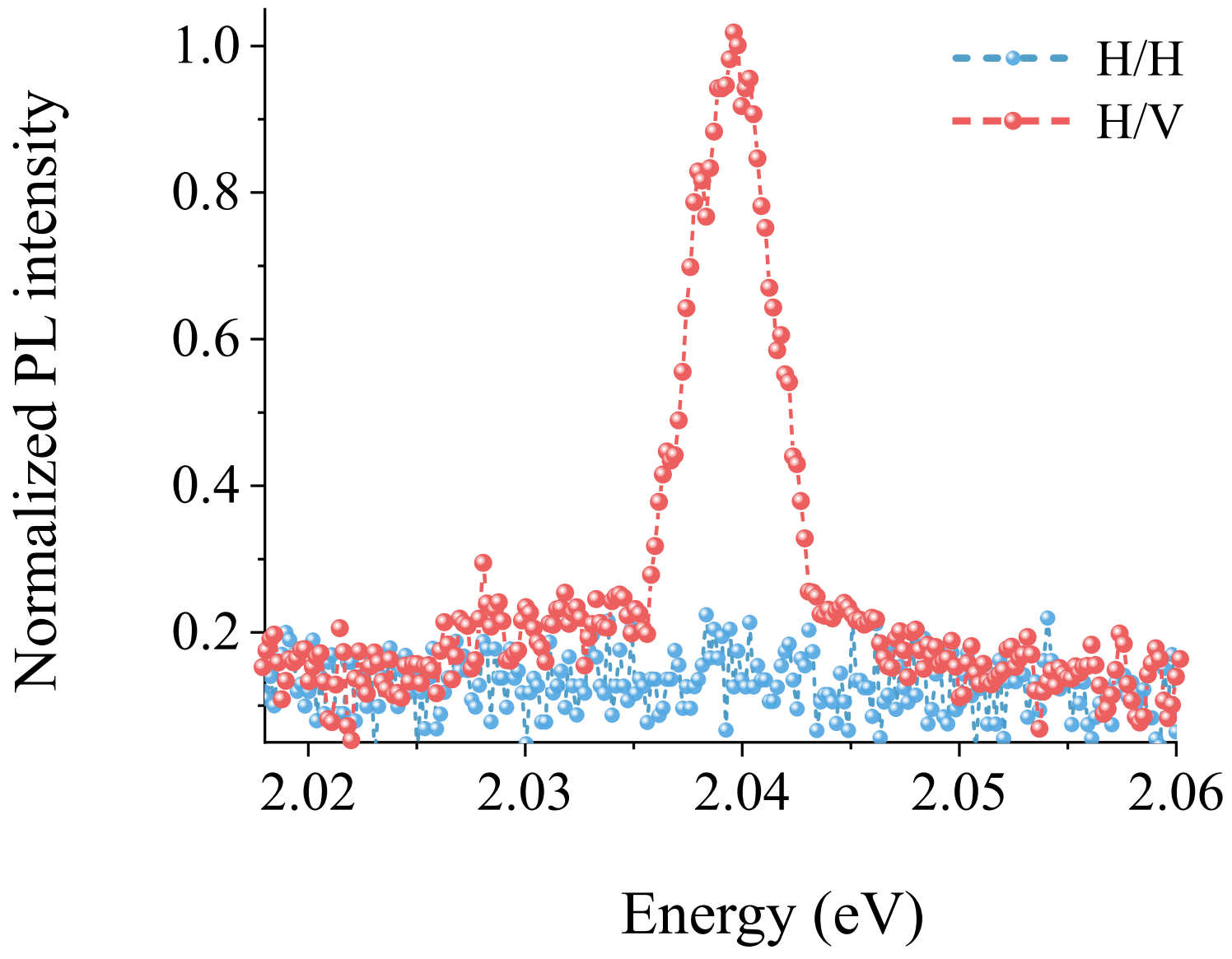

**Supplementary Figure 7a:** The co- (H/H, in red) and cross- (H/V, in blue) linearly polarized PL spectra taken on the few-layer graphene encapsulated $WS_2$ (GWG) sample showing ~100% DOLP at 5 K. The excitation wavelength is 593 nm.

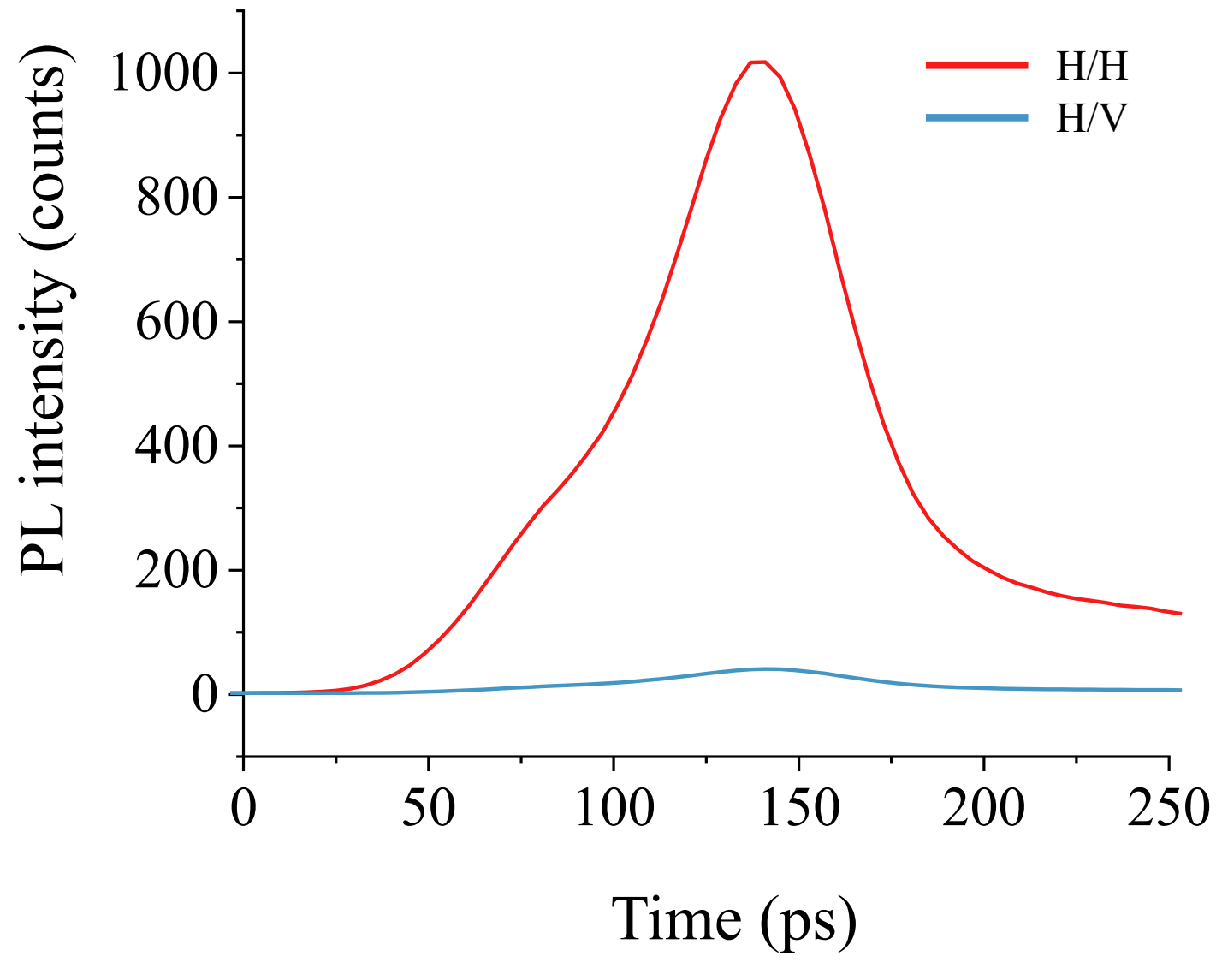

**Supplementary Figure 7b:** Polarization Resolved TRPL measurement showing a peak DOLP of 97.6% in the GWG stack at 5 K obtained with near-resonant excitation at 593 nm.

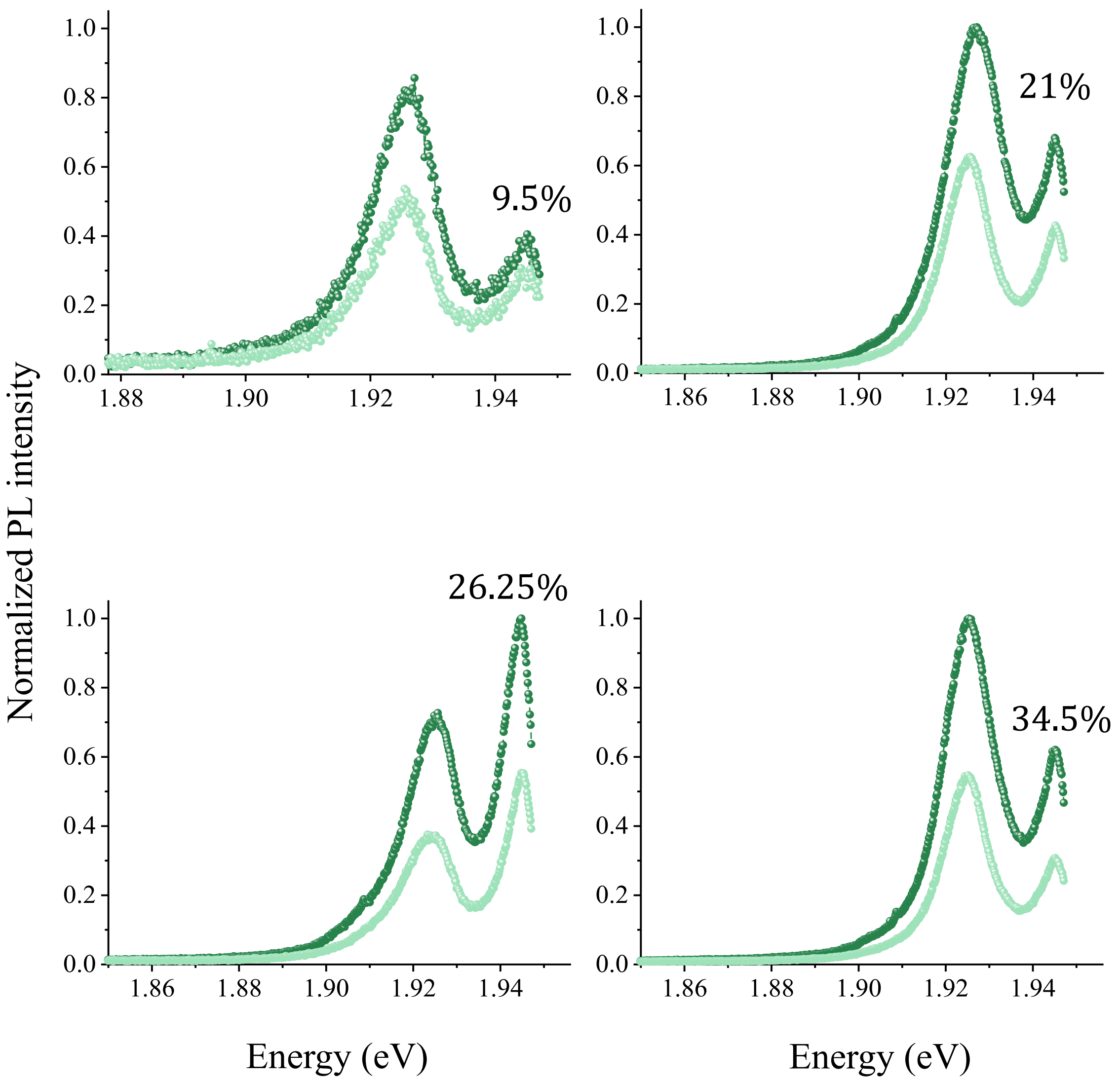

**Supplementary Figure 8:** The co- ($\sigma+/\sigma+$, in dark green) and cross- ($\sigma+/\sigma-$, in light green) circularly polarized PL spectra obtained at different spots on the HMH stack.

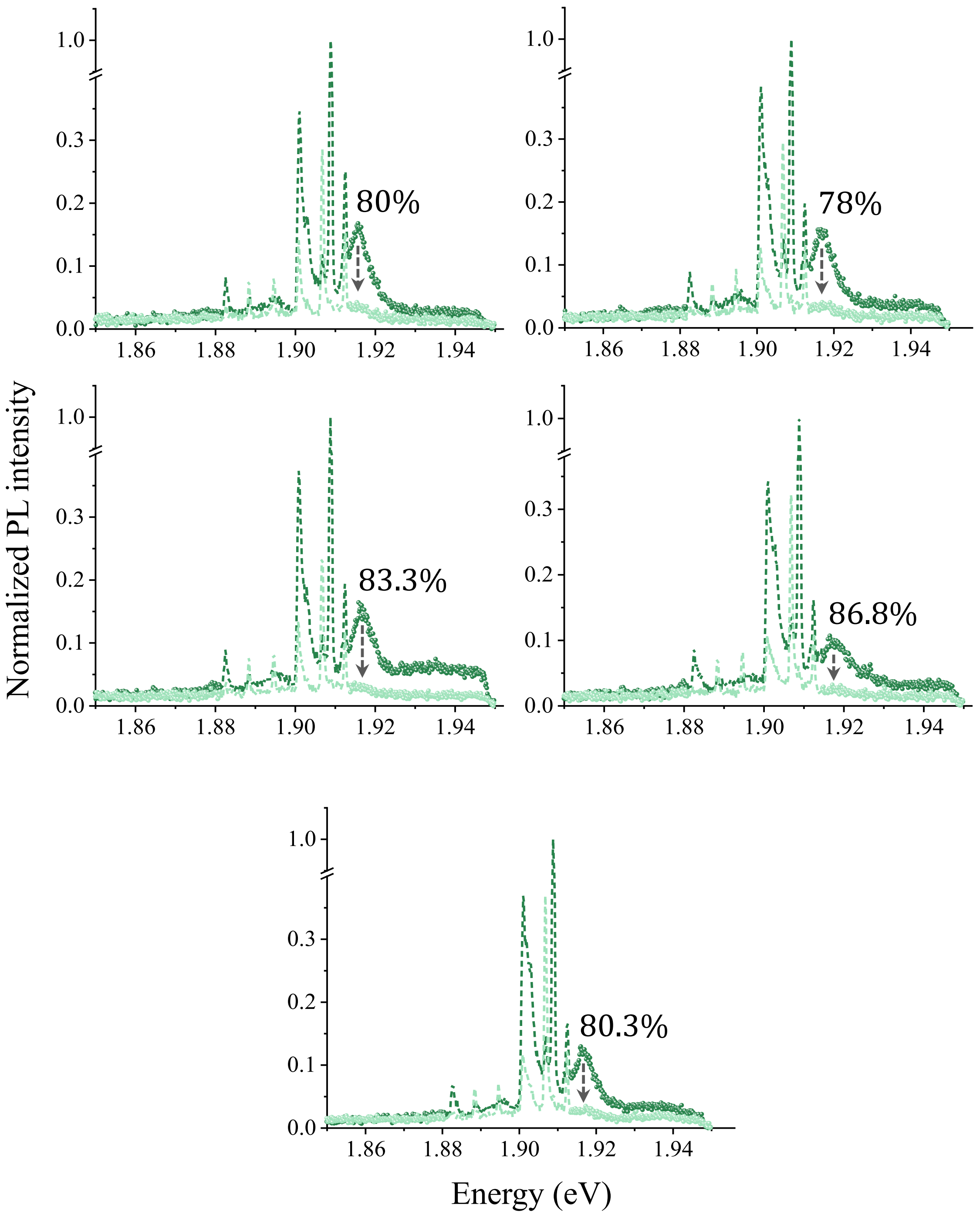

**Supplementary Figure 9:** The co- ($\sigma+/\sigma+$, in dark green) and cross- ($\sigma+/\sigma-$, in light green) circularly polarized PL spectra obtained at different spots on the GMG stack. The peaks indicated by the dashed lines represent Raman peaks. The dashed lines are the strong Raman peaks due to dual resonance in the GMG stack.

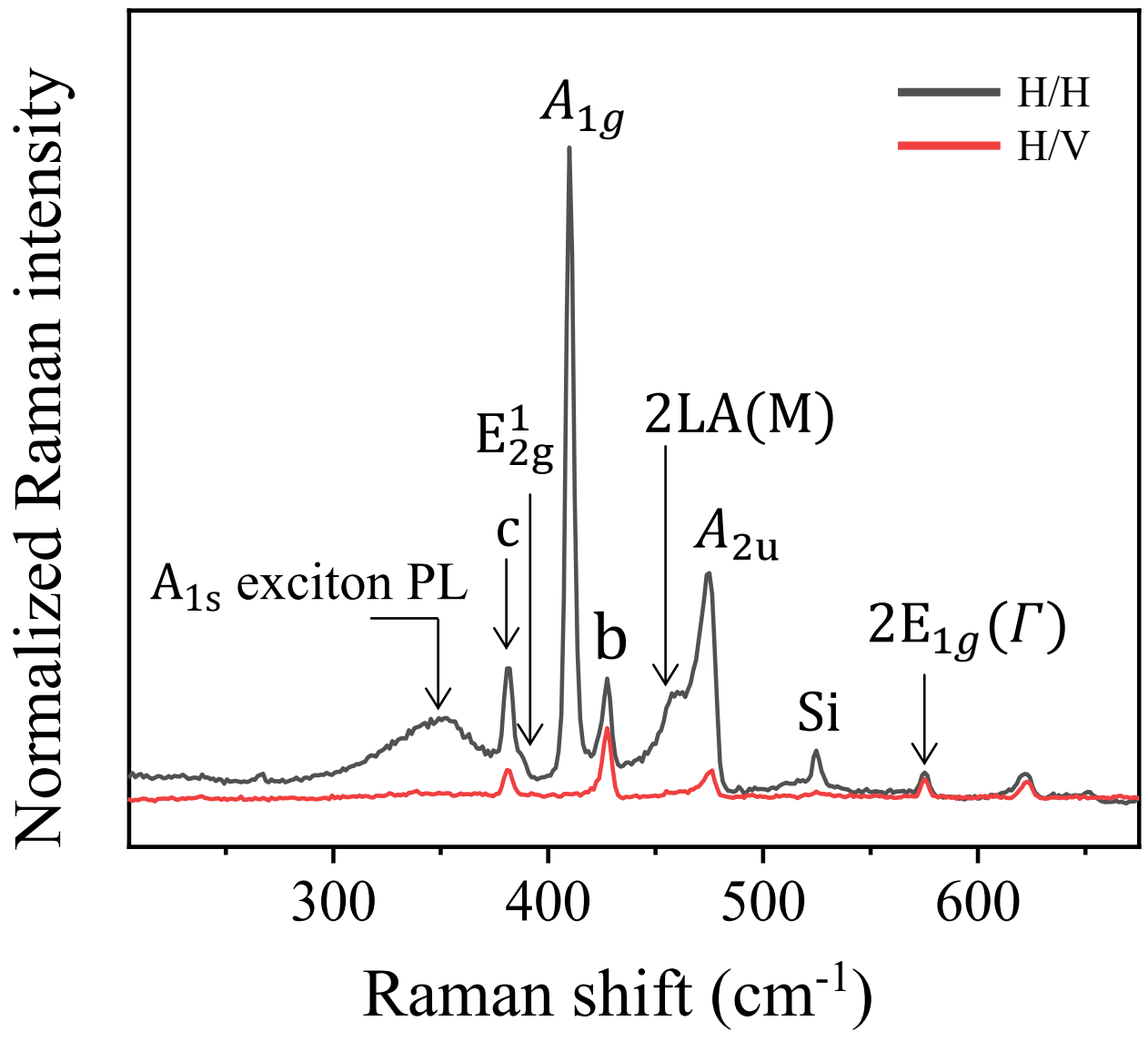

**Supplementary Figure 10:** The linear polarization resolved Raman spectra of the GMG stack on 633 nm near-resonant excitation at 5 K. The nomenclature of the Raman peaks is adopted from Chakraborty et al.[8]

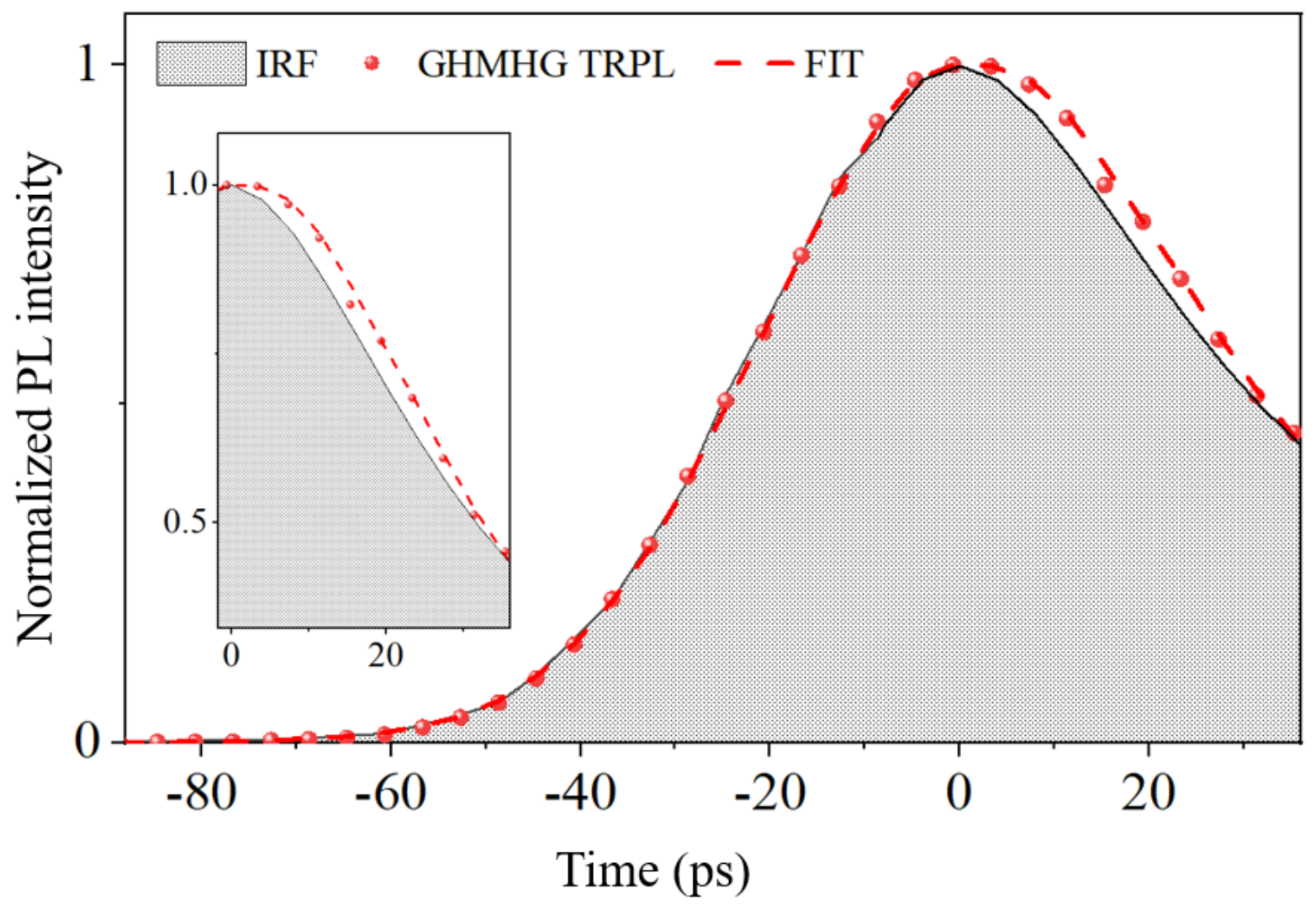

**Supplementary Figure 11:** The time-resolved PL of the $A_{1s}$ exciton in monolayer $MoS_2$ in the GHMHG stack taken at 5 K. The red symbols (red dashed trace) indicate the measured data (fitting). The instrument response function (IRF) is indicated by the grey shaded region. Inset: The zoomed-in view of the decay part in log scale.

We use the following procedure to deconvolute the IRF contribution and to extract the sample response. The response $[L(t)]$ of the system to an impulse excitation is given as follows:

$$L(t) = \sum_i A_i e^{-t/\tau_i}$$

Here $A_i$ and $\tau_i$ represent the amplitude and time constant of the $i^{th}$ contributing species in the TRPL response, respectively. The above impulse response is convoluted with the experimentally obtained instrument response function (IRF) $[I(t)]$ (shaded region) iteratively, ensuring that the final convoluted output (dashed line) precisely matches the experimentally obtained TRPL data $[T(t)]$ (red symbols), i.e.,

$$T(t) = \int_0^t I(t')L(t-t')dt'$$

The number in the inset of the figure above represents the fastest component in the decay indicating the exciton lifetime at one of the spots.

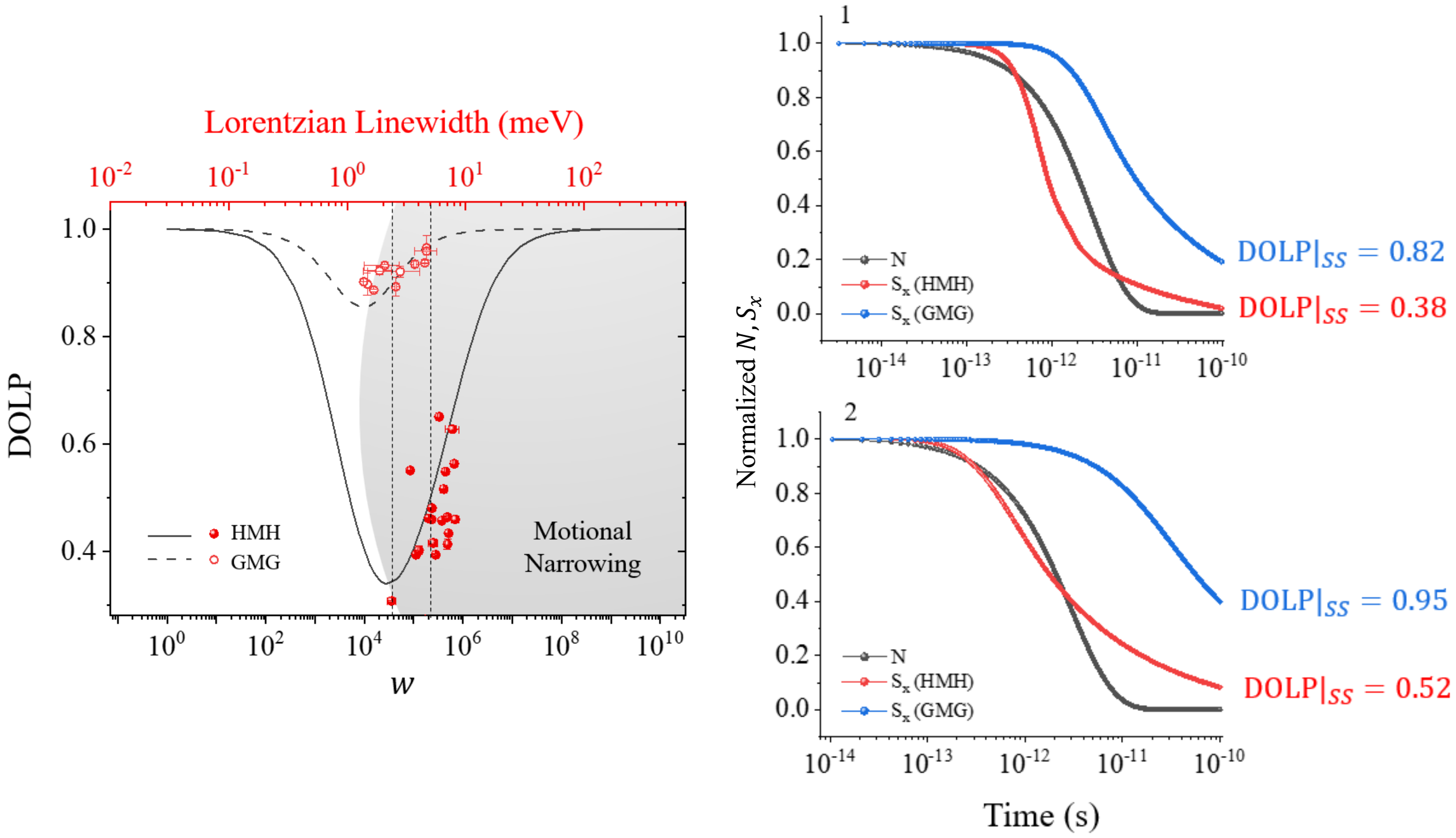

**Supplementary Figure 12:** Left panel - Simulation results of the steady-state MSS equation (lines) and experimental (symbols) results comparing the exciton DOLP as a function of $w$ (bottom axis) and Lorentzian linewidth (top axis) [Figure 4 in the main text]. The shaded region indicates motional narrowing. Right panel – Simulation results of the time-dependent MSS equation for the HMH (in red) and the GMG (in blue) stack. The top and the bottom plot corresponds to two different values of $w$, as marked by dashed vertical lines in the left panel. To verify the simulation results, we estimate the steady-state DOLP value using the following expression:

$$\mathrm{DOLP}|_{SS} = \frac{\sum_t \mathrm{N(t)}\, S_x(\mathrm{t})}{\sum_t \mathrm{N(t)}}$$

The steady-state DOLP value for the HMH and the GMG stack is also shown in the above plot and is in good agreement with the experimentally obtained DOLP values at that $w$. The time-dependent results of the MSS equation clearly show the giant enhancement of the valley coherence time in the GMG stack as compared to the HMH stack.